\documentclass[aps,prc,twocolumn,superscriptaddress,nofootinbib]{revtex4-1}

\usepackage{dcolumn}
\usepackage{graphicx,color,subfigure}
\usepackage{textcomp}
\usepackage{multirow}
\usepackage{relsize}
\definecolor{bobcatgreen}{rgb}{0.05,0.31,0.25}

\begin{document}
  \newcommand {\nc} {\newcommand}
  \nc {\beq} {\begin{eqnarray}}
  \nc {\eeq} {\nonumber \end{eqnarray}}
  \nc {\eeqn}[1] {\label {#1} \end{eqnarray}}
  \nc {\eol} {\nonumber \\}
  \nc {\eoln}[1] {\label {#1} \\}
  \nc {\ve} [1] {\mbox{\boldmath $#1$}}
  \nc {\ves} [1] {\mbox{\boldmath ${\scriptstyle #1}$}}
  \nc {\mrm} [1] {\mathrm{#1}}
  \nc {\half} {\mbox{$\frac{1}{2}$}}
  \nc {\thal} {\mbox{$\frac{3}{2}$}}
  \nc {\fial} {\mbox{$\frac{5}{2}$}}
  \nc {\la} {\mbox{$\langle$}}
  \nc {\ra} {\mbox{$\rangle$}}
  \nc {\etal} {{\it et al.}}
  \nc {\eq} [1] {(\ref{#1})}
  \nc {\Eq} [1] {Eq.~(\ref{#1})}
  \nc {\Refc} [2] {Refs.~\cite[#1]{#2}}
  \nc {\Sec} [1] {Sec.~\ref{#1}}
  \nc {\chap} [1] {Chapter~\ref{#1}}
  \nc {\anx} [1] {Appendix~\ref{#1}}
  \nc {\tbl} [1] {Table~\ref{#1}}
  \nc {\Fig} [1] {Fig.~\ref{#1}}
  \nc {\ex} [1] {$^{#1}$}
  \nc {\Sch} {Schr\"odinger }
  \nc {\flim} [2] {\mathop{\longrightarrow}\limits_{{#1}\rightarrow{#2}}}
  \nc {\textdegr}{$^{\circ}$}
  \nc {\inred} [1]{\textcolor{red}{#1}}
  \nc {\inblue} [1]{\textcolor{blue}{#1}}
  \nc {\IR} [1]{\textcolor{red}{#1}}
  \nc {\IB} [1]{\textcolor{blue}{#1}}
  \nc{\pderiv}[2]{\cfrac{\partial #1}{\partial #2}}
  \nc{\deriv}[2]{\cfrac{d#1}{d#2}}

  \nc{\ai}{{\it ab~initio}}
  \nc{\Ai}{{\it Ab~initio}}
  


\def\lsim{\mathrel{\rlap{\lower4pt\hbox{\hskip1pt$\sim$}}
    \raise1pt\hbox{$<$}}}         
\def\gsim{\mathrel {\rlap{\lower4pt\hbox{\hskip1pt$\sim$}}
    \raise1pt\hbox{$>$}}}         
    
\title{Dissecting reaction calculations using Halo EFT and {\it ab initio} input}

\author{P. Capel}
\email[]{pcapel@uni-mainz.de}
\affiliation{Institut f\"ur Kernphysik, Johannes Gutenberg-Universit\"at Mainz, 55099 Mainz, Germany}
\affiliation{Physique Nucl\' eaire et Physique Quantique (CP 229), Universit\'e libre de Bruxelles (ULB), B-1050 Brussels}
\affiliation{Institut f\"ur Kernphysik, Technische Universit\"at Darmstadt, 64289 Darmstadt, Germany}
\affiliation{ExtreMe Matter Institute EMMI, GSI Helmholtzzentrum f{\"u}r Schwerionenforschung GmbH, 64291 Darmstadt, Germany}

\author{D.~R. Phillips}
\email[]{phillid1@ohio.edu}
\affiliation{Institute of Nuclear and Particle Physics and Department of Physics and Astronomy, Ohio University, Athens, OH 45701,USA}
\affiliation{Institut f\"ur Kernphysik, Technische Universit\"at Darmstadt, 64289 Darmstadt, Germany}
\affiliation{ExtreMe Matter Institute EMMI, GSI Helmholtzzentrum f{\"u}r Schwerionenforschung GmbH, 64291 Darmstadt, Germany}

\author{H.-W. Hammer}
\email[]{hans-werner.hammer@physik.tu-darmstadt.de}
\affiliation{Institut f\"ur Kernphysik, Technische Universit\"at Darmstadt, 64289 Darmstadt, Germany}
\affiliation{ExtreMe Matter Institute EMMI, GSI Helmholtzzentrum f{\"u}r Schwerionenforschung GmbH, 64291 Darmstadt, Germany}

\date{\today}

\begin{abstract}
  We present a description of the break-up of halo nuclei in peripheral nuclear reactions by coupling a model of the projectile motivated by Halo Effective Field Theory with a fully dynamical treatment of the reaction using the Dynamical Eikonal Approximation.
  Our description of the halo system reproduces its long-range properties, i.e., binding energy and asymptotic normalization coefficients of bound states and phase shifts of continuum states.
  As an application we consider the break-up of $^{11}$Be in collisions on Pb and C targets. Taking the input for our Halo-EFT-inspired description of $^{11}$Be from a recent \ai\ calculation of that system yields a good description of the Coulomb-dominated breakup
  on Pb at energies up to about 2 MeV, with the result essentially independent of the short-distance part of the halo wave function.
  However, the nuclear dominated break-up on C is more sensitive to
  short-range physics. The role of spectroscopic factors and possible
  extensions of our approach to include additional short-range mechanisms are
  also discussed.
 \end{abstract}

\pacs{24.10.−i, 24.87.+y, 25.60.Gc, 25.60.Tv, 25.70.De}
\keywords{Exotic nuclei, Coulomb breakup, nuclear breakup, }

\maketitle

\section{\label{intro}Introduction}

The quantitative description of  nuclear structure and reactions
on the same footing is a major challenge of contemporary nuclear
theory~\cite{BC12,Thompson:2014vda,Cizewski:2015uqa}.
\Ai\ approaches to calculate nuclear scattering
observables are limited by the computational complexity of the nuclear
many-body problem. This limitation applies especially to exotic isotopes
along the neutron and proton drip lines which are weakly bound or unbound.
With new radioactive beam facilities such as FRIB and FAIR
on the horizon, the
quest for improved approaches for nuclear reactions with exotic
isotopes has become a major topic in the nuclear-theory community.
The ultimate goal of this effort is to have a robust and reliable
model of nuclei and nuclear reactions with predictive power and
quantified uncertainties~\cite{Carlson:2017ebk}.

In this work we focus on the structure and reactions of halo nuclei.
Halo nuclei are weakly-bound objects consisting of one or more valence
nucleons and a tightly-bound core
nucleus~(see, e.g., Refs.~\cite{Jonson:2004,Riisager:2012it}). They
exemplify the emergence of new effective degrees of freedom close to the
drip lines.
Accurate models for the breakup of halo nuclei have been shown to be sensitive mostly to the tail of the wave function \cite{CN07} for both the bound state and the continuum states \cite{CN06}.
This means that the structure observables to which these calculations are sensitive are the one-neutron separation energy and the Asymptotic Normalization Coefficient (ANC) for the bound state and the phase shifts in the continuum.
This suggests that it is not really necessary to include
a detailed (and computationally expensive) microscopic description of a
halo projectile in reaction models.
On the contrary, an effective two-body description of the projectile which replicates the experimental information and/or the results of a microscopic nuclear-structure model for on-shell quantities, like ANCs and phase shifts, to a given accuracy should be enough to obtain a reaction model with good predictive power.

Here we consider the example of the one-neutron halo nucleus \ex{11}Be.
  \ex{11}Be has two bound levels which can be viewed as a neutron and a \ex{10}Be core in a relative $s$- and $p$-wave, respectively.
For \ex{11}Be the ANCs and \ex{10}Be-n scattering phase shifts were recently obtained in an \ai\ No-Core Shell Model with Continuum (NCSMC) calculation~\cite{CNR16}.
Although it needed to be tuned phenomenologically to correctly reproduce the experimental binding energies, this calculation is the most thorough extant microscopic description of \ex{11}Be.

In this paper, we take a first step towards the ultimate goal expressed in Ref.~\cite{Carlson:2017ebk} by 
complementing the Dynamical Eikonal Approximation (DEA) from Refs.~\cite{BCG05,GBC06}	
with the expansion of the Halo Effective Field Theory (Halo-EFT) for \ex{11}Be, developed in
Ref.~\cite{Hammer:2011ye}. 
Halo EFT is based on an expansion in powers of the distance scale associated with the \ex{10}Be core, $R_{\rm core}$, divided by that associated with the \ex{11}Be halo, $R_{\rm halo}$ (see Ref.~\cite{Hammer:2017tjm} for a recent review of Halo-EFT).
Consideration of the relative sizes of these nuclei, together with the results of
Ref.~\cite{Hammer:2011ye} suggests an expansion parameter of $R_{\rm core}/R_{\rm halo}\sim0.4$. 
Reference \cite{Hammer:2011ye} used Halo-EFT to compute the differential E1 strength of \ex{11}Be; when combined with a simplified reaction model this reproduces data on the Coulomb dissociation of ${}^{11}$Be on a ${}^{208}$Pb target~\cite{Palit:2003av}.

Our aim here is to use a sophisticated reaction theory to describe this collision, while also identifying the nuclear-structure inputs---which can be obtained from \ai\ calculations or directly from experiment---needed for a description of the breakup reaction at a given accuracy.
The amplitudes computed in Ref.~\cite{Hammer:2011ye} are not suitable for incorporation in a sophisticated reaction theory. Such a theory of the breakup of neutron halos on a target requires the specification of target-core, target-neutron, and core-neutron interactions. Therefore in this work we develop a series of Halo-EFT potentials to simulate the \ex{10}Be-$n$ interaction.
Instead of pure contact interactions, we use for practical reasons potentials with a Gaussian shape and ranges
1.7, 2.1, and 2.8 fm---all of which are of the order of the distance scale $R_{\rm core}$.

We construct versions of these potentials at leading order (LO) and next-to-leading order (NLO) in the Halo-EFT calculation of the E1-dissociation strength of \ex{11}Be in Ref.~\cite{Hammer:2011ye}.\footnote{We note that the NLO and higher-order potentials will be iterated to all orders, in contrast to the strictly perturbative treatment in Halo EFT. We will discuss this in detail in Sec.~\ref{sec:iteration}.}
We also construct a third set of potentials, in which the NLO EFT interaction is supplemented by potentials in the $p_{3/2}$, $d_{5/2}$, and $d_{3/2}$ channels that reproduce known \ex{11}Be resonances. This allows us to diagnose the ingredients necessary to reproduce features observed in reaction cross sections at higher energies.  This last ``NLO EFT + resonances" set of potentials can be regarded as a proxy for a Halo EFT of \ex{11}Be in which additional degrees of freedom corresponding to those resonances are included. We use a combination of experimental data (e.g., for bound-state and resonance energies) and \ai\ results (e.g., for ANCs and phase shifts) to determine the pertinent potential parameters. 

Comparing the results obtained with these three sets of Halo-EFT potentials therefore allows us to trace which on-shell \ex{10}Be-$n$ matrix elements affect which features of the breakup cross section.
The use of different ranges in the Gaussian potential provides various interiors of the \ex{10}Be-$n$ wave function while keeping the on-shell properties fixed.
This enables us to estimate the influence of short-distance physics in the reaction mechanism.

The paper is structured as follows. We begin by describing the ingredients of our calculation. In Sec.~\ref{dea}, we describe the reaction theory used to compute the collision of the \ex{11}Be projectile with the lead and carbon targets. 
Then, in Sec.~\ref{eft}, we summarize the key features, power counting, and results of the Halo-EFT treatment of \ex{11}Be from Ref.~\cite{Hammer:2011ye}.
Section~\ref{abinitio} summarizes the \ai\ calculation~\cite{CNR16} that provides the input we use to fix the potentials.
The EFT expansion encourages a focus on interactions in the $s_{1/2}$ and $p_{1/2}$, and so we describe the construction of potentials for those two channels in Sec.~\ref{potentials1}.  
Results of the reaction calculations for \ex{11}Be impinging on \ex{208}Pb and $^{12}$C targets at about 70~MeV/nucleon using those potentials are presented and compared to experiment \cite{Fuk04} in Sec.~\ref{results1}.
Section~\ref{potentials2} revisits the construction of potentials, explaining how the interactions in the resonant $p_{3/2}$, $d_{5/2}$, and $d_{3/2}$ channels are constructed, and Sec.~\ref{results2} shows results for \ex{11}Be breakup with these ``NLO + resonances" potentials.
Finally, Sec.~\ref{unobservable} provides some examples of quantities to which the breakup cross section is not sensitive and Sec.~\ref{conclusion} offers our conclusions. 
 
\section{\label{dea}Three-cluster description of the collision}
To describe the collision of a one-neutron halo nucleus, like \ex{11}Be, on a target, such as Pb or C, we consider the usual few-body framework of reaction theory (see Ref.~\cite{BC12} for a recent review).
The projectile $P$ is seen as a core $c$---$^{10}$Be in the present case---to which a valence neutron $n$ is loosely bound.
This two-cluster structure is described by the effective Hamiltonian
\begin{equation}
  H_0=-\frac{\hbar^2\nabla_r^2}{2 \mu} + V(\ve{r})
  \label{eq:Hprojectile}
\end{equation}
where $\ve{r}$ is the $c$-$n$ relative coordinate and $\mu=m_c m_n/m_P$ is the $c$-$n$ reduced mass, with $m_c$, $m_n$, and $m_P=m_c+m_n$ the masses of the core, the neutron and the projectile, respectively.
The potential $V$ simulates the interaction between the halo neutron and the core.
Its explicit form is described in Secs.~\ref{potentials1} and \ref{potentials2} below.
In this simple effective description of the halo nucleus, the internal structure of the core is neglected and its spin is assumed to be zero.

The relative motion between the halo neutron and the core is described by the eigenstates of $H_0$.
In the partial wave of orbital angular momentum $l$ and total angular momentum $J$, which is obtained from the composition of $l$ and the neutron spin $s$, and projection $M$, they read
\beq
H_0\varphi_{lJM}(E,\ve{r})=E\ \varphi_{lJM}(E,\ve{r}),
\eeqn{e0}
with
\beq
\varphi_{lJM}(E,\ve{r})=\frac{u_{lJ}(E,r)}{r}\left[\chi_s\otimes Y_l(\hat{\ve{r}})\right]^{JM},
\eeqn{e1}
where $\chi_s^{m_s}$ are spinors and $Y_l^m$ are spherical harmonics.

Negative-energy states [$E<0$ in \Eq{e0}] are discrete and correspond to the spectrum of the projectile below the one-neutron separation threshold.
We add the number of nodes $n$ in the radial wave function to the other quantum numbers to enumerate them.
The reduced radial wave functions of these bound states exhibit the following asymptotic behavior
\beq
u_{nlJ}(E_{nlJ},r)\flim{r}{\infty}{\cal C}_{nlJ}\ \kappa_{nlJ} r\ k_l(\kappa_{nlJ}r),
\eeqn{eANC}
where $k_l$ is a modified spherical Bessel function of the second kind, $\kappa_{nlJ}=\sqrt{2\mu |E_{nlJ}|/\hbar^2}$, and ${\cal C}_{nlJ}$ is the asymptotic normalisation coefficient (ANC) associated with that bound state.
Note that for $s$-waves the combination $x k_0(x)=e^{-x}$.

The positive-energy states [$E>0$ in \Eq{e0}] describe the continuum of the nucleus.
They exhibit the following asymptotic behavior
\beq
u_{lJ}(E,r)\flim{r}{\infty}\cos\delta_{lJ}(E)\, kr\, j_{l}(kr)+\sin\delta_{lJ}(E)\, kr\, n_{l}(kr),\nonumber\\
\eeqn{edep}
where $j_l$ and $n_l$ are the regular and irregular spherical Bessel functions, respectively, $k=\sqrt{2\mu E/\hbar^2}$ is the wave number for the $c$-$n$ relative motion, and $\delta_{lJ}$ is the phase shift. 

In the center-of-mass frame of the $P$-$T$ system our three-body model of the collision is specified by the Hamiltonian
\beq
H=-\frac{\hbar^2\nabla_R^2}{2 \mu_{PT}}+H_0+U_{cT}(R_{cT})+U_{nT}(R_{nT}).
\eeqn{e2}
Here $\ve{R}$ is the coordinate of the projectile center of mass relative to the target and $\mu_{PT}=m_P m_T/(m_P+m_T)$ is the $P$-$T$ reduced mass, with $m_T$ the target mass.
The optical potentials $U_{cT}$ and $U_{nT}$ simulate the $c$-$T$ and $n$-$T$ interactions, respectively.
They depend on the $c$-$T$ and $n$-$T$ relative distances $R_{cT}$ and $R_{nT}$, respectively.
We neglect the internal structure of the target $T$, and use optical potentials from the literature for $U_{cT}$ and $U_{nT}$.

Within this framework, studying the collision reduces to solving the three-body Schr\"odinger equation
\beq
H\, \Psi(\ve{r},\ve{R}) = {\cal E}\, \Psi(\ve{r},\ve{R}),
\eeqn{e3}
where $\cal E$ is the total energy in the $P$-$T$ center-of-mass rest frame and the wave function $\Psi$ describes the thee-body relative motion.
Equation \eq{e3} has to be solved with the condition that the projectile, initially in its ground state $\varphi_{n_0l_0J_0M_0}$ of energy $E_{n_0l_0J_0}$, is impinging on the target
\beq
\Psi^{(M_0)}(\ve{r},\ve{R})\flim{Z}{-\infty}e^{i{\cal K}Z}\, \varphi_{n_0l_0J_0M_0}(\ve{r}),
\eeqn{e4}
where the $Z$ axis has been taken along the beam direction.
In \Eq{e4}, the projectile-target initial momentum $\hbar{\cal K}$ is related to the total energy ${\cal E}=\hbar^2{\cal K}^2/2\mu_{PT}+E_{n_0l_0J_0}$.

Various methods have been developed to solve that problem (see Ref.~\cite{BC12} for a recent review).
In the time-dependent approach (TD), the $P$-$T$ relative motion is simulated by a classical trajectory, while the internal structure of the projectile is treated quantum mechanically \cite{KYS94,EBB95,TS01r,CBM03c}.
It is affected by a time-dependent field because of its interaction with the target.
This leads to the resolution of a time-dependent \Sch equation.
On the contrary, the coupled-channels technique with a discretized continuum (CDCC) is a fully quantal approach, within which the three-body wave function $\Psi$ is expanded upon the projectile eigenstates $\varphi_{lJM}(E)$ \cite{Kam86,Aus87,TNT01}.
To be numerically tractable, the $c$-$n$ continuum is discretized into energy ``bins''.
This expansion leads to a set of coupled equations.
Finally, the eikonal approximation assumes that, at sufficiently high energy, most of the $\ve{R}$ dependence of $\Psi$ is in the initial plane wave $e^{i{\cal K}Z}$ of \Eq{e4} \cite{HT03,OYI03,BCG05}.
Factorizing this plane wave out of $\Psi$ simplifies the three-body \Sch equation \eq{e3}.
Additional simplification can be obtained by performing a subsequent adiabatic approximation \cite{HT03}.

In the present study, we choose to use the Dynamical Eikonal Approximation (DEA) \cite{BCG05,GBC06}, which, although based on the eikonal approximation, does not include the adiabatic treatment of the projectile.
This model works very well for one-neutron \cite{GBC06} and one-proton \cite{GCB07} halo nuclei impinging on both light and heavy targets. 
The DEA has been compared to the TD and CDCC methods in Ref.~\cite{CEN12}.
It exhibits similar computational times to the TD approach but naturally includes quantal interferences, which are absent in the TD method due to its inherent semiclassical hypothesis.
At the intermediate energies considered here, the DEA shows excellent agreement with the fully quantal CDCC calculations in different breakup observables (viz. energy and angular distributions).
We therefore believe the results presented below do not depend on the particulars of the reaction model and can be seen as general.

\section{\label{eft} Halo EFT for Coulomb dissociation of \ex{11}Be}

The nucleus $^{11}$Be is the archetypical one-neutron halo nucleus.
Its valence neutron is bound by a mere $S_{1n}= 503$~keV \cite{AJZENBERGSELOVE19901}\footnote{\label{n1}Note that there is a newer value
$S_{1n}= 501.6$ keV for the neutron separation energy of $^{11}$Be~\cite{KKP12} but this 0.3\% change is far below the accuracy of our calculation.}, and its $\frac{1}{2}^+$ ground state corresponds predominantly to a $1s_{1/2}$ neutron bound to a $^{10}$Be core in its $0^+$ ground state.
Thanks to this loose binding and the absence of any centrifugal or Coulomb barrier, the valence nucleon can tunnel far into the classically-forbidden region and hence has a high probability of being a large distance from the core.
At least qualitatively, this explains the large one-neutron removal and breakup cross sections observed experimentally for \ex{11}Be \cite{Tan96}.

This ``intruder" ground state differs from the $\frac{1}{2}^-$ expected in an extreme shell model, which predicts a $0p_{1/2}$ valence neutron structure.
That $p$ state corresponds to the first---and only---bound excited state of \ex{11}Be: it lies 184~keV below the one-neutron threshold.
Above the neutron threshold, various states can be interpreted as single-particle resonances, including the $\frac{5}{2}^+$ state at $E=1.274$~MeV in the continuum, which is often seen as a $d_{5/2}$ resonant neutron coupled to the $0^+$ ground state of the $^{10}$Be core \cite{Fuk04,CGB04}.

Halo EFT provides a systematic treatment of (bound or unbound) nuclei in which, as in $^{11}$Be, the last few nucleons are loosely bound compared to a nuclear core (see Ref.~\cite{Hammer:2017tjm} for a recent review).
In Halo EFT, quantum-mechanical amplitudes are expanded in powers of the small parameter $R_{\rm core}/R_{\rm halo}$, where $R_{\rm core}$ ($R_{\rm halo}$) is the size of the core (halo) of the nucleus.
The degrees of freedom are then the halo nucleons and the core.
These are treated as structureless at leading order in the expansion, although structure is included through higher-order terms, much as in a multipole expansion.
The breakdown scale of the theory is then set by the size of the core---reactions that resolve its features will not be properly described in the theory---or by the energy required to excite the core---reactions that proceed through the excitation of the core also cannot be described in this most basic version of the theory.

The theory can be extended by including ``core excitation", at the cost of introducing additional parameters~\cite{Zhang:2013kja,Zhang:2014zsa,Ryberg:2014exa}.
In the case of our \ex{11}Be calculation we do not include the $2^+$ state of \ex{10}Be core at $E(2^+)=3.4$ MeV above the ground state as an explicit degree of freedom.
Thanks to the loose binding of the valence neutron in $^{11}$Be, we obtain an expansion parameter of
$\sqrt{S_{1n}/E(2^+)}\approx 0.4$.
A similar estimate is obtained from the ratio of sizes of the core to the halo.

Halo EFT is  formulated through a Lagrangian that includes all operators up to a given order in this expansion.
The interactions that appear in this Lagrangian are contact interactions and derivatives thereof, and so correspond to zero-range potentials.
The coefficients of these operators, the ``low-energy constants" (LECs) of the theory, are free parameters, that must be adjusted to reproduce experimental data or inputs from nuclear-structure calculations.
At leading order (LO) there is only one interaction, which appears in the $s_{1/2}$ channel. Its coefficient is tuned to reproduce $S_{1n}=0.503$ MeV.
At next-to-leading order (NLO) another parameter appears in this channel.
For reactions involving the \ex{11}Be bound state it is optimal to tune this parameter to the ANC of the ground state, ${\cal C}_{1s1/2}$ [see \Eq{eANC}] \cite{Phillips:1999hh}.
Up to the order we consider this is equivalent to fixing the $s_{1/2}$ effective range, $r_{s1/2}$.
The parameters in this channel then scale as $S_{1n} \sim (2 \mu R_{\rm halo}^2)^{-1}$ and $r_{s1/2} \sim R_{\rm core}$. As a consequence the ANC scales as
\begin{equation}
{\cal C}_{1s1/2}=\sqrt{\frac{2 \kappa_{1s1/2}}{1-r_{s1/2} \kappa_{1s1/2}}} \sim R_{\rm halo}^{-1/2}.
\end{equation}

If $p$-wave interactions are ``natural", i.e. of the order anticipated by naive dimensional analysis with respect to the scale $R_{\rm core}$~\cite{Bedaque:2003wa}, then these are the only free parameters in the Coulomb dissociation amplitude up to $\mathcal{O}[(R_{\rm core}/R_{\rm halo})^3]$ in the expansion~\cite{Hammer:2011ye}.
This situation is realized for ${}^{19}$C, where a good description of the data of Ref.~\cite{Nakamura:1999rp} can be obtained with only $S_{1n}$ and ${\cal C}_{1s1/2}$ as inputs~\cite{Acharya:2013nia}.
It is also the situation in the $p_{3/2}$ channel of the \ex{10}Be-$n$ system, since there is no resonance or bound-state with an energy $\lsim 1$ MeV there.
The corresponding expression for the breakup spectrum induced by the E1 operator is~\cite{Hammer:2011ye}
\begin{widetext}
\begin{equation}
\frac{d{\rm B(E1)}}{dE}^{(J=3/2),  {\rm NLO}}=
\frac{4\, Q_{\rm eff}^2 e^2 \mu}{\hbar^2 \pi^2} {\cal C}_{1s1/2}^2 \frac{k^3}{(k^2 +\kappa^2_{1s1/2})^4},
\label{eq:dBdE1NLOp32}
\end{equation}
\end{widetext}
where $Q_{\rm eff}=Z_c\, m_n/m_P \approx 4/11$ for \ex{11}Be.
This $\frac{d {\rm B(E1)}}{dE}$ is a universal function, in the sense that the E1 dissociation spectrum for any weakly-bound system has the same shape (up to N$^3$LO corrections) as long as there are no enhanced final-state $p$-wave interactions~\cite{Typel:2004zm,TB05,Bertulani:2009zk,Hammer:2017tjm}. 

However, in the $p_{1/2}$ channel such interactions {\it are} present, since they must be strong enough to generate the $\frac{1}{2}^-$ \ex{11}Be excited bound state, whose one-neutron separation energy is $S^*_{1n}=184$~keV~\cite{AJZENBERGSELOVE19901}.
As already shown by Typel and Baur~\cite{Typel:2004zm,TB05}, those $p$-wave interactions significantly affect the Coulomb dissociation spectrum.
Two parameters are needed to describe the low-energy physics at LO in the $p_{1/2}$ channel: both $S^*_{1n}$, and the $p$-wave effective ``range", $r_{p1/2}$, are required as input to the LO EFT amplitude in this channel.
The necessity to fix two parameters for a LO description of a $p$-wave bound state, in contrast to the sole parameter needed for a LO description of an $s$-wave bound state, results from the need to properly renormalize the propagator that describes this state~\cite{Bertulani:2002sz}.
The parameters scale as $S_{1n}^* \sim (2 \mu R_{\rm halo}^2)^{-1}$, $r_{p1/2} \sim R_{\rm core}^{-1}$.  We note that the $p$-wave effective range determines ${\cal C}_{0p1/2}$, the ANC of the $p$-wave bound state, at LO in the Halo EFT expansion~\cite{Hammer:2011ye}.

Since the Coulomb dissociation experiment is not one where the E1 transition accesses the $p$-wave bound state (cf. Ref.~\cite{Kwan:2014dha}) the $p_{1/2}$ channel is entered ``off resonance". Consequently, the $p_{1/2}$ amplitude that encodes final-state interactions can be expanded in powers of $R_{\rm core}/R_{\rm halo}$~\cite{Hammer:2011ye}. The NLO result for the E1 dissociation spectrum in this channel is then:
\begin{widetext}
\beq
\frac{d{\rm B(E1)}}{dE}^{(j=1/2),  {\rm NLO}}&=&
\frac{2\, Q_{\rm eff}^2 e^2 \mu}{\hbar^2 \pi^2}
\frac{k^3}{(k^2 + \kappa_{1s1/2}^2)^4} \left({\cal C}_{1s1/2}^2
+ \frac{4 \kappa_{1s1/2}^2}{r_{p1/2}} \frac{\kappa_{1s1/2}^2  + 3 k^2}{k^2  + \kappa^2_{0p1/2}}
\right) \, .
\eeqn{eq:dBdE1NLOp12}
\end{widetext}
The second term in the round brackets includes final-state interactions and is formally NLO: it is suppressed by $R_{\rm core}/R_{\rm halo}$ relative to the first term. Since $r_{p1/2} < 0$ it acts to reduce the cross section.
Equation (\ref{eq:dBdE1NLOp12}) agrees with the scaling formulae derived in Refs.~\cite{Typel:2004zm,TB05} if we take $k^3 \cot \delta_{p1/2}=\frac{r_{p1/2}}{2} (k^2 + \kappa_{0p1/2}^2)$---in accordance with the power counting for a shallow $p$-wave bound state~\cite{Bedaque:2003wa,Hammer:2017tjm}---and neglect terms of $O(R^2_{\rm core}/R_{\rm halo}^2)$.
The total $\frac{d {\rm B(E1)}}{dE}$ for \ex{11}Be is then obtained by summing (\ref{eq:dBdE1NLOp32}) and (\ref{eq:dBdE1NLOp12}). 

The parameters $S_{1n}$ and $S_{1n}^*$ are well determined, but Ref.~\cite{Hammer:2011ye} did not thoroughly explore the parameter choices for ${\cal C}_{1s1/2}$ and $r_{p1/2}$---or, equivalently, ${\cal C}_{0p1/2}$---, i.e., how to determine the ANCs of the two bound states of \ex{11}Be.
These values have not been directly measured.
Fortunately, the work of Calci {\it et al.}~\cite{CNR16} provides numbers for these two inputs from an \ai\ eleven-body calculation of this system. 

Once these two parameters are fixed the Halo EFT of \ex{11}Be predicts the distribution of E1 strength with energy (and with angle for that matter, see Ref.~\cite{Acharya:2013nia}).
It also predicts the B(E1) value for the $\frac{1}{2}^+ \rightarrow \frac{1}{2}^-$ transition, as well as the electric radius of the ground state.
These quantities agree with experiment for the parameter values chosen in Ref.~\cite{Hammer:2011ye}.
In that paper predictions were also made for the---as yet unmeasured---electric radius of the $\frac{1}{2}^-$ state. 

Finally, we note that integrating out the $2^+$ core excitation effectively generates a short-range interaction between the target and \ex{11}Be in the Coulomb break-up reaction.
This interaction is small at momenta significantly below $1/R_{\rm core}$.
It cannot be included at the level of the reduced transition strength discussed in this section but would naturally appear in a full reaction calculation as discussed below.
In the present work, however, such terms are not considered.

\section{\label{abinitio} Ab initio calculation of \ex{11}Be by Calci {\it et al.} \cite{CNR16}}

Due to the strong clustering observed in their structure and the intrinsically large extension of their wave function, halo nuclei are real challenges for \ai\ nuclear-structure calculations.
In the recent Ref.~\cite{CNR16}, Calci \etal\ have performed a {\it tour de force} by successfully computing the nuclear structure of $^{11}$Be \ai\ within the ``No-core Shell Model with Continuum'' (NCSMC) \cite{Baroni13L,Baroni:2013fe,Navratil:2016ycn}.

The no-core shell model (NCSM) performs an exact diagonalization of a particular $A$-body Hamiltonian that can include two- and three-nucleon forces within a harmonic-oscillator basis~\cite{Navratil:2007we}.
This model converges well for nuclei whose size is of order the oscillator length.
Hence systems like \ex{11}Be that are weakly bound converge very slowly with basis size in the NCSM.
The solution to this is to include in the model space additional basis functions that have the cluster structure observed in the halo system.
If the clusters contain $a$ and $b$ particles respectively (with $A=a+b$) then these clustered wave functions are products of solutions of the Hamiltonian in the $a$- and $b$-body spaces, with the wave function representing the relative degree of freedom between the two clusters initially unknown~\cite{Baroni13L,Baroni:2013fe,Navratil:2016ycn}:
\beq
\lefteqn{|\Psi (J^\pi) \rangle=\sum_\lambda |A \lambda J^\pi \rangle}\nonumber\\
 &+& \sum_{{\rm channels} \; ab,\nu} \frac{\gamma_\nu(r_{ab})}{r_{ab}} {\cal A}_\nu [|a \lambda_a J^\pi_a \rangle |b \lambda_b J^\pi_b \rangle]^{s} Y_{l}(\hat{r}_{ab})]^{J^\pi}.
\eeqn{eq:NCSMC}
The first term of this expression corresponds to the standard NCSM expansion of the $A$-body wave function, with $\lambda$ indicating the quantum numbers of the harmonic-oscillator basis (orbital-angular momentum, spins\ldots).
The second term includes all relevant two-body clusterizations and has $\nu$ as a collective label that enumerates the states in the product representation of the two harmonic oscillator bases, coupled to appropriate spins ($J_a$ and $J_b$ for each of the clusters, coupled to a total spin $s$), relative orbital angular momentum $l$, total angular momentum $J$, and isospin (not shown): $\nu=\{\lambda_a J^\pi_a; \lambda_b J^\pi_b;s,l,J\}$.
The wave function describing the relative motion between both clusters is also a function of the relative coordinate between the centers of mass of the clusters, $\ve{r}_{ab}$.
The antisymmetrizer ${\cal A}_\nu$ ensures that the resulting wave function is antisymmetric. (For a full discussion with more explicit notation, see Refs.~\cite{Baroni:2013fe,Navratil:2016ycn}.) 

The basis (\ref{eq:NCSMC}) is over-complete, but this is dealt with using an extension of the standard RGM norm-kernel procedure~\cite{Navratil:2016ycn}.
The clustered part of the NCSMC basis contains exactly the states needed to improve the convergence of the diagonalization for halos built on those clusters.
Scattering phase shifts are also accessible via matching the wave function to the scattering state: in practice this is done via an $R$-matrix technique. 

In their calculation of $^{11}$Be, Calci \etal\ considered a family of chiral EFT forces including two- and three-nucleon interactions \cite{CNR16}.
They found that without including the continuum, i.e., when only the NCSM is considered (the first term in Eq.~\eq{eq:NCSMC}), no converged $^{11}$Be can be obtained, confirming the significant clusterization of its structure.
Even with the basis extended as in Eq.~(\ref{eq:NCSMC}), the shell inversion between the bound states is difficult to achieve, and only the $\rm N^2LO_{sat}$ NN force reproduces it.
That interaction is obtained by adjusting simultaneously the two- and three-nucleon interactions to reproduce low-energy nucleon-nucleon scattering data, as well as binding energies and radii of few-nucleon systems and selected isotopes of carbon and oxygen \cite{Ekstrom:2015rta}.

Since reaction calculations are quite sensitive to the exact value of the energies of the states, Calci \etal\ have phenomenologically tuned some matrix elements of the interaction in the \ex{11}Be system to ensure these energies are correctly reproduced.
These NCSMC-pheno calculations then provide predictions for nuclear-structure properties such as the bound-state ANCs (see Tables~\ref{t1} and \ref{t2}) \cite{CNR16}.
As mentioned in \Sec{eft}, these predictions help us constrain the parameters of the effective potential we use to simulate the $^{10}$Be-$n$ interaction within our reaction calculations (see \Sec{potentials1}).

In addition to improving the discrete spectrum of the nucleus, the inclusion of the continuum in this \ai\ model provides predictions for the phase shift in the $^{10}$Be-$n$ continuum, in both resonant (e.g., $5/2^+$) and non-resonant (e.g., $1/2^+$) partial waves.
These predictions first help us check the quality of the fits performed on the bound states (see  Secs.~\ref{s1} and \ref{p1}).
Second, they help us constrain the $^{10}$Be-$n$ interaction in higher partial waves (see \Sec{potentials2}).

\section{\label{potentials1}Construction of \ex{10}Be-$n$ potentials}
\label{Gaussians}
\subsection{Potential shape}

We now turn our attention to the Hamiltonian that describes the \ex{10}Be-$n$ system, $H_0$,
given in Eq.~(\ref{eq:Hprojectile}).
Following the philosophy of EFT, we exploit the fact that low-energy dynamics in this system should not be sensitive to details of the short-distance physics.
As explained in \Sec{eft}, this suggests that contact interactions plus derivatives of contact interactions can be used for the $^{10}$Be-$n$ potential $V$ in \Eq{eq:Hprojectile}.
The Halo EFT Lagrangian is formulated as a derivative expansion and the operators that appear at order $2l$ in the derivative expansion can be mapped to yield potentials that are different in each $(lJ)$ channel.
A potential derived from Halo EFT can thus be fitted partial wave
by partial wave, similar to the treatment of the short-range part of the
chiral EFT potential in Ref.~\cite{Epelbaum:1999dj}. We exploit this fact to construct our Halo EFT potential on the partial-wave basis.
The Halo EFT  for ${}^{11}$Be described in Sec.~\ref{eft} implies that at NLO the potential in both $s_{1/2}$ and $p_{1/2}$ waves includes a contact term and an operator corresponding to two derivatives of a contact term, while the potential in all other partial waves is zero.
To evaluate the sensitivity of the reaction to the short-range physics of the projectile, and to render the $s_{1/2}$ and $p_{1/2}$ interactions numerically tractable, we regulate them with a Gaussian, whose range can be varied. 
To simplify its handling, the potential is parametrized as follows
\beq
V_{lJ}(r)=V^{(0)}_{lJ}\ e^{-\frac{r^2}{2\sigma^2}}+V^{(2)}_{lJ}\ r^2e^{-\frac{r^2}{2\sigma^2}}.
\eeqn{e5}
This potential provides us with two adjustable parameters per partial wave, viz. the depths $V^{(0)}_{lJ}$ and $V^{(2)}_{lJ}$.
The fact that the parameters are adjusted in each partial wave is notated in Eq.~(\ref{e5}) by the subscript $lJ$ added to the potential and its fitting parameters.
As will be detailed in the following sections, they are fitted so as to reproduce the asymptotic properties of the projectile that affect the reaction calculation, viz. the binding energies of the two \ex{11}Be bound states and their ANCs, or the phase shifts in other prominent partial waves~\cite{CN07,CN06}.
By constructing the potential \eq{e5} in this way we reproduce observables in the partial waves where the potential is active and ensure there is no interaction in the other waves.
When possible, we use experimental data for this fitting.
However, since there is no direct experimental determination of the ANC of the $^{11}$Be bound states nor of the $^{10}$Be-$n$ phase shifts, we also rely on the \ai\ calculation of Calci \etal\ to fully determine the LECs of the potential \eq{e5}. 

The range of the Gaussian $\sigma$ is an unfitted parameter that is varied to evaluate the sensitivity of the reaction calculations to this effective description of the projectile.
We consider $\sigma=1.2$, 1.5, and 2~fm. We define the range of the potential
as the distance scale at which the potential has decreased to $e^{-1}\approx
0.368$ of its value at the origin.
In terms of momentum, this corresponds to the scale
$\Lambda \sim 1/(\sqrt{2}\sigma)$. For momenta of order
$\Lambda$ and larger (or distances below $1/\Lambda$), the
details of the potential matter and the effective theory breaks down.
We do not consider potentials with $\sigma$ larger than 2.0 fm, since such large co-ordinate space cutoffs would produce distortions of the long-distance physics of the \ex{11}Be system. The smallest $\sigma$ considered, 1.2 fm, is already markedly smaller than the 
size of the \ex{10}Be nucleus, which---as explained above---can be taken as the distance scale at which this description of \ex{11}Be breaks down. 

We note that simple Gaussian potentials like these have recently been used to describe the nucleon-nucleon interaction~\cite{Kievsky:2016kzb}. In that case, in the three-nucleon and four-nucleon systems the Gaussian nucleon-nucleon interaction must be supplemented by a three-nucleon interaction (for which a Gaussian regulator may also be chosen). We emphasize that we do not expect that the potentials constructed here could, on their own, describe Beryllium isotopes further along the isotopic chain.  Any study of ${}^{12}$Be that builds on the ideas laid out here would have to supplement the ${}^{10}$Be-$n$ potential by a ``three-body" ${}^{10}$Be-$n$-$n$ interaction that was tuned to reproduce a low-energy property of ${}^{12}$Be. Using three-body forces to redress the deficiencies of simple two-body potentials in bigger systems is very much in the spirit of Halo and pionless EFT~\cite{Bedaque:1998kg,Bedaque:1998km,Bedaque:1999ve,Canham:2008jd,Rotureau:2012yu,Ji:2014wta}.

\subsection{\label{s1}$s_{1/2}$ wave}

In the $s_{1/2}$ partial wave, the two potential parameters are adjusted to reproduce the one-neutron separation energy in the ground state of $^{11}$Be and its ANC.
For the former, we choose the experimental value $S_{1n}=503$~keV cited in Ref.~\cite{AJZENBERGSELOVE19901}, while for the latter, we use the prediction of Calci \etal\ \cite{CNR16}.
The corresponding depths are listed in \tbl{t1} with the actual energy $E_{1s1/2}$ and ANC  ${\cal C}_{1s1/2}$ obtained for each potential identified by its width $\sigma$.
The scattering lengths $a_{s1/2}$ and effective ranges $r_{s1/2}$ obtained with these potentials are also provided.
The last line provides the results of the NCSMC calculation \cite{CNR16}.

\begin{table}
\begin{tabular}{c|cc|cccc}\hline\hline
$\sigma$ & $V^{(0)}_{s1/2}$ & $V^{(2)}_{s1/2}$ & $E_{1s1/2}$ & ${\cal C}_{1s1/2}$ & $a_{s1/2}$ & $r_{s1/2}$\\ 
(fm) & (MeV) & (MeV fm\ex{-2}) & (MeV) &  (fm\ex{-1/2}) & (fm) & (fm)\\\hline
1.2 & $-50.375$ & $-45$ & $-0.5031$ & 0.7865 & 9.20 & 3.72\\
1.5 & $-100.19$ & 0 & $-0.5033$ & 0.791 & 9.23 & 3.76\\
2 & $-80.755$ & $+3$ & $-0.5031$ & 0.7845 & 9.21 & 3.76\\\hline
\ai & - & -& $-0.5$ & 0.786 & 9.21 & 3.68\\
\hline\hline
\end{tabular}
\caption{\label{t1} Depths of the Gaussian \ex{10}Be-$n$ potential \eq{e5} in the $s_{1/2}$ partial wave.
The energy and ANC of the $1s_{1/2}$ state obtained through this fit are also provided, as well as the scattering length $a_{s1/2}$ and effective range $r_{s1/2}$.
The \ai\  predictions of Calci \etal\ are listed in the last line \cite{CNR16}.}
\end{table}

\begin{figure}[ht]
\includegraphics[width=\linewidth]{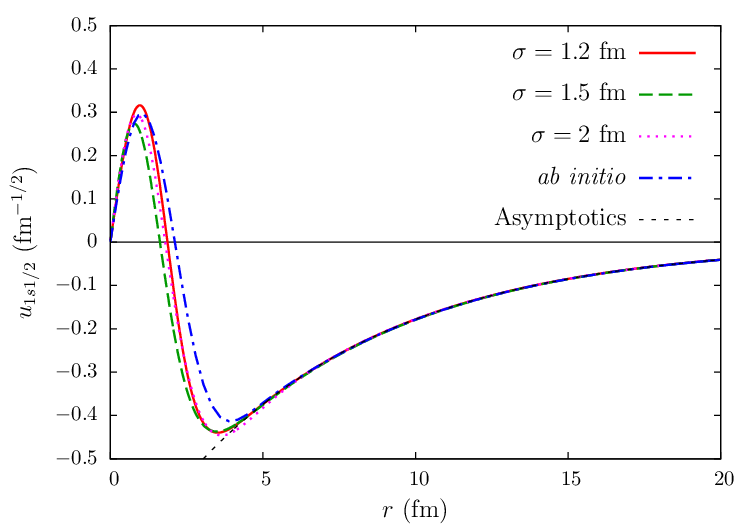}
\caption{\label{f1} Radial wave functions for the \ex{11}Be $\half^+$ ground state obtained within the effective description of the nucleus.
The \ex{10}Be-$n$ overlap wave function obtained from the NCSMC calculation of Calci \etal\  \cite{CNR16} is shown as well.}
\end{figure}

The reduced radial wave functions thereby obtained for this $1s_{1/2}$ state are displayed in \Fig{f1} for the three potential ranges considered in this study: $\sigma=1.2$~fm (red solid line), 1.5~fm (green dashed line), and 2~fm (magenta dotted line).
Because of the way they were constructed, these three potentials lead to the same asymptotics of the wave function.
However, they produce different behaviors in its interior.
The \ai\ overlap wave function is shown as well for comparison (blue dash-dotted line).
By construction, the wave function of the potentials agree with it at large distances, but they start to deviate from it as soon as the \ai\ overlap function ceases to follow the asymptotic form shown by the thin dashed line.

\begin{figure}[ht]
\includegraphics[width=\linewidth]{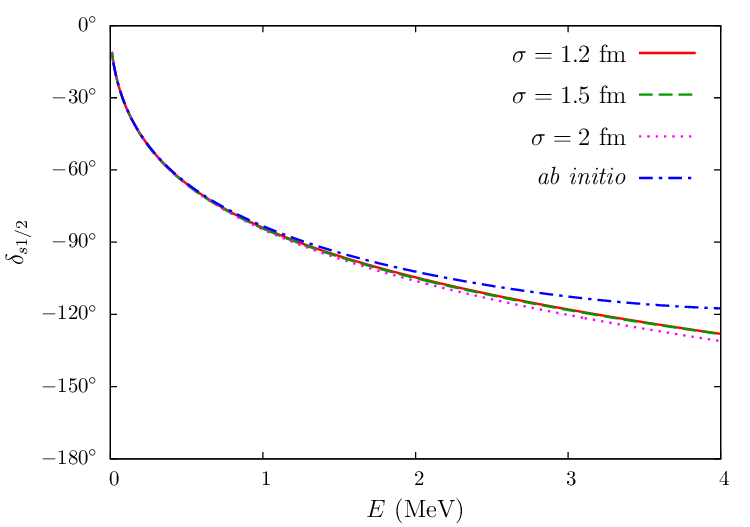}
\caption{\label{f2} Phase shifts obtained in the $s_{1/2}$ partial wave as a function of the \ex{10}Be-$n$ relative energy $E$.
The phase shift predicted by the NCSMC calculation is shown as well \cite{CNR16}.}
\end{figure}

Because breakup calculations are sensitive not only to the initial ground state of the projectile, but also to the description of its final states, i.e. its continuum \cite{CN06}, we examine in \Fig{f2} the $s_{1/2}$ phase shift generated by the three potentials adjusted to the ground-state properties and compare them to the NCSMC results.
Effective-range theory guarantees that, for a shallow bound state like this one, there will be a tight correlation between $E_{1s1/2}$ and ${\cal C}_{1s1/2}$, on the one hand, and the scattering length and effective range on the other~\cite{SCB10}.
Accordingly, the scattering length and effective range for these three potentials are very similar (see Table~\ref{t1}).
Interestingly, they are also in excellent agreement with the \ai\ ones.
Therefore, all three potentials provide nearly identical phase shifts, which agree perfectly with the \ai\ phase shift up to about 1.5 MeV.

\subsection{\label{p1}$p_{1/2}$ wave}

In the $p_{1/2}$ wave we fit the potential parameters in a similar way using data for the $\half^-$ excited bound state of \ex{11}Be: $S_{1n}^*=184$~keV \cite{AJZENBERGSELOVE19901} and the ANC ${\cal C}_{0p1/2}$ obtained by Calci \etal\ \cite{CNR16,Cal17}.
The fitted depths of the Gaussian potentials in that partial wave and the nuclear-structure outputs obtained from these values are listed in \tbl{t2} along with the predictions of the NCSMC calculation of Calci \etal\ \cite{CNR16}.

\begin{table}
\begin{tabular}{c|cc|cccc}\hline\hline
$\sigma$ & $V^{(0)}_{p1/2}$ & $V^{(2)}_{p1/2}$ & $E_{0p1/2}$ & ${\cal C}_{0p1/2}$ & $a_{p1/2}$ & $r_{p1/2}$\\ 
(fm) & (MeV) & (MeV fm\ex{-2}) & (MeV) &  (fm\ex{-1/2}) & (fm\ex{3}) & (fm\ex{-1})\\\hline
1.2 & $-96.956$ & 0 & $-0.1841$ & 0.1288 & 235 & $-1.23$\\
1.5 & $-83.625$ & $+5.2$ & $-0.1842$ & 0.1290 & 236 & $-1.22$\\
2.0 & $-57.504$ & $+3.3$ & $-0.1841$ & 0.1295 & 243 & $-1.17$\\\hline
\ai & - & -& -0.1848 & 0.1291 & 237 & $-1.21$\\
\hline\hline
\end{tabular}
\caption{\label{t2} Depths of the Gaussian \ex{10}Be-$n$ potential \eq{e5} 
in the $p_{1/2}$ partial wave.
The energy and ANC of the $0p_{1/2}$ state obtained by this fit are provided, as well as the scattering ``length'' $a_{p1/2}$ and effective ``range'' $r_{p1/2}$.
The results of the NCSMC calculation are listed in the last line \cite{CNR16,Cal17}.}
\end{table}

\begin{figure}[ht]
\includegraphics[width=\linewidth]{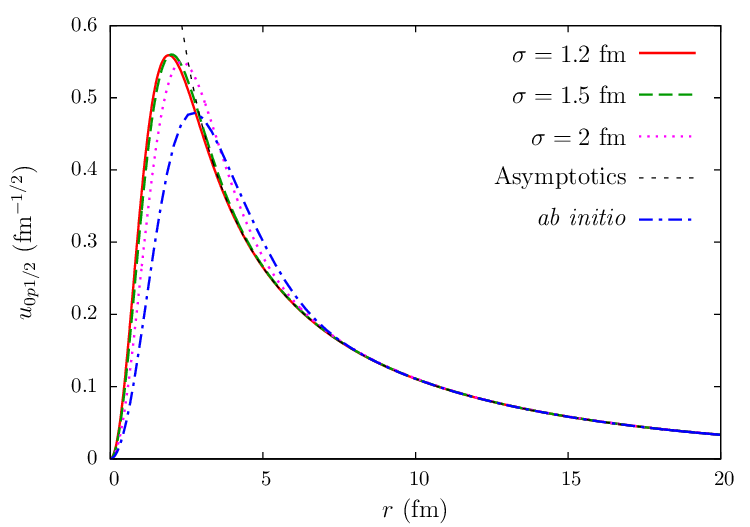}
\caption{\label{f3} Radial wave functions for the \ex{11}Be $\half^-$ excited state obtained within the effective description of the nucleus.
The \ex{10}Be-$n$ overlap wave function obtained from the NCSMC calculation of Calci \etal\ is shown as well \cite{CNR16,Cal17}.}
\end{figure}

The corresponding $0p_{1/2}$ radial wave functions are displayed in \Fig{f3}, alongside the \ai\ overlap wave function.
We observe this time that, although all wave functions exhibit the same ANC, 
they do not reach their asymptotic behavior at the same radius.
While the wave functions obtained with the narrow potentials ($\sigma=1.2$~fm and 1.5~fm) reach their asymptotics at a rather short distance ($r<5$~fm) as in the $1s_{1/2}$ state, the broad potential ($\sigma=2$~fm) does not reach it before $r\sim 7$~fm.
Interestingly, this is also observed for the \ai\ prediction.
This suggests that the couplings with other channels in the NCSMC calculation of Calci \etal\ extend to large distances and that they are attractive at the surface of the nucleus.
Since this state corresponds to the expected shell-model state, significant couplings with various configurations are expected, in particular with those in which the $^{10}$Be core is in one of its excited states \cite{Nav17}.
This might explain the late reach of the asymptotics by the NCSMC overlap wave function for this state.
However this has not been investigated by Calci \etal\ in Ref.~\cite{CNR16}.

\begin{figure}[ht]
\includegraphics[width=\linewidth]{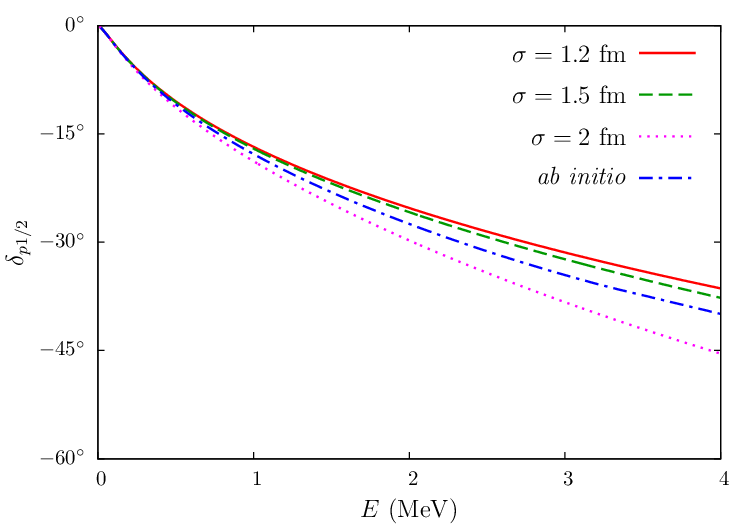}
\caption{\label{f4} Phase shifts obtained in the $p_{1/2}$ partial wave as a function of the \ex{10}Be-$n$ relative energy $E$.
The NCSMC calculation is shown as well \cite{CNR16}.}
\end{figure}

The $p_{1/2}$ phase shifts obtained by the three Gaussian potentials exhibit similar energy dependences, especially at low energy (see \Fig{f4}).
This follows from the fact that they have  very similar effective-range expansion parameters (see \tbl{t2}).
However the differences between them are larger than in the $s_{1/2}$ wave, a phenomenon that is probably related to the non-zero orbital angular momentum \cite{SCB10}.
In particular, the $\sigma=2$~fm potential leads to a phase shift that is significantly lower than the others at $E\gsim1$~MeV.
Nevertheless, all three potentials provide $p_{1/2}$ phase shifts that are in fair agreement with the \ai\ prediction on the whole energy range \cite{CNR16}.

\subsection{Leading-order potentials}\label{LO}

In order to estimate the significance of the NLO terms in the description of \ex{11}Be on the reaction dynamics, we also construct a LO potential.
At LO, the only parameter in the Halo EFT calculation of E1 breakup of \ex{11}Be is the one-neutron separation energy of the ground state \cite{Hammer:2017tjm}.
To reproduce this observable, we take only the first term of the effective potential \eq{e5} in the $s_{1/2}$ partial wave and set the potential to zero in all other partial waves.
In particular, the potential in the $p_{1/2}$ channel is taken to be zero, in accord with the above Halo-EFT calculation, where the $p_{1/2}$-wave interaction's effect on the reaction cross section only appears at NLO (see \Sec{eft}).
We adjust our sole LO parameter to the one-neutron separation energy of \ex{11}Be.
The ANC is then {\it not} controlled, and some variation in the asymptotic part of these potential-model wave functions results.
The fitted depths and the energy and ANC of the resulting $1s_{1/2}$ state are displayed in \tbl{t1LO}.

\begin{table}
\begin{tabular}{c|c|cc}\hline\hline
$\sigma$ & $V^{\rm LO}_{s1/2}$ & $E_{1s1/2}$ & ${\cal C}^{\rm LO}_{1s1/2}$\\ 
(fm) & (MeV) & (MeV) &  (fm\ex{-1/2}) \\\hline
1.2 & -153.25 & -0.5032 & 0.735\\
1.5 & -100.19 & -0.5033 & 0.791\\
2 & -58.40 & -0.5031 & 0.892\\
\hline\hline
\end{tabular}
\caption{\label{t1LO} Depths of the LO (purely central) Gaussian \ex{10}Be-$n$ potential that serves to simulate the $\half^+$ ground state of \ex{11}Be.
The energy and ANC of the resulting $1s_{1/2}$ state are also provided.}
\end{table}

\subsection{Iteration of higher orders and Wigner bound}
\label{sec:iteration}
We note that for practical reasons higher-order interactions will be
iterated in our approach. Strictly applying the power counting of Halo EFT
(cf.~Sec.~\ref{eft} and Ref.~\cite{Hammer:2017tjm}), however, 
such higher-order interactions should be treated in perturbation theory.
Two comments are in order here.
First, such a partial inclusion of higher-order terms will not make the
calculation more accurate since some higher-order contributions are
missing.
Second, the iteration of higher-order interactions should not
create any problems as the additionally included contributions are
small.

However, closer inspection reveals that the second argument is too simplistic.
The iteration of higher-order terms can alter the ultraviolet behavior
of the amplitude and generate additional divergences which cannot be
renormalized by the terms present in the effective interaction
to the given order. Such a partial resummation is 
permissible if the regularization cutoff $\sqrt{2}\sigma\equiv 1/\Lambda$
is kept above or close to the breakdown scale $R_{\rm core}$, since then
the unrenormalized terms are of natural size~\cite{Lepage:1997cs}.
This partial resummation approach has, e.g.,
been used for two-neutron halos and the triton
\cite{Canham:2009xg,Bedaque:2002yg}.

This restriction placed on the regulator is connected to the Wigner
causality bound \cite{Wigner:1955zz} 
which constrains the minimum range of short-range (energy-independent)
interactions if exact unitarity is to be maintained. 
The corresponding constraints on EFT interactions have been worked out
in Refs.~\cite{Phillips:1996ae,Hammer:2009zh,Hammer:2010fw}.
In our case, the Wigner bound corresponds to a constraint on the values of the
parameter $\sigma$ in Eq.~(\ref{e5}) and effectively excludes values below
1 fm.

\section{\label{results1} Breakup of \ex{11}Be on Pb and C at 70~MeV/nucleon with Halo-EFT potentials}

\subsection{Two-body inputs}

In this section, we initiate our study of the influence of the effective description of the halo nucleus on reaction observables by analyzing the breakup of \ex{11}Be into \ex{10}Be+$n$ on Pb at 69~MeV/nucleon and C at 67~MeV/nucleon.
These reactions have been measured at RIKEN by Fukuda \etal\ \cite{Fuk04} and various breakup observables have been obtained.\footnote{Note that we refrain from discussing the GSI data by Palit \etal\ \cite{Palit:2003av} since they are at a much higher beam energy of 520~MeV/nucleon where relativistic effects may be important. A study dedicated to these effects is in progress \cite{MC18}.}
They therefore constitute excellent test cases for our study.

In addition to the description of the \ex{11}Be projectile, for which we use the Halo-EFT potentials described in \Sec{potentials1}, the few-body model of the reaction requires optical potentials to simulate the interaction between the projectile constituents and the target.
The potentials are the same as those used in Refs.~\cite{CBM03c,CGB04}

In the collision on the Pb target, we use the Becchetti and Greenlees parametrization for the $n$-Pb optical potential \cite{BG69}.
The \ex{10}Be-Pb interaction is simulated by an $\alpha$-Pb optical potential \cite{Bon85} scaled to account for the different size of the nuclei.
Since,  at this energy, this reaction is dominated by Coulomb excitation of \ex{11}Be the particular choice of optical potential does not make a large difference to the final results \cite{CBM03c}.

When we consider the collision on C, we follow Ref.~\cite{CGB04} for the choice of the optical potentials.
For the $n$-\ex{12}C interaction, we use a potential developed by Comfort and Karp that reproduces the elastic scattering of protons on \ex{12}C at energies between 12 and 183~MeV \cite{CK80}.
To obtain a potential at the energy of interest, we use a linear interpolation between the parameters obtained at 61.4~MeV and 96~MeV.
The \ex{10}Be-C interaction is simulated by the optical potential developed by Al-Khalili, Tostevin and Brooke to reproduce the experimental elastic-scattering cross section of \ex{10}Be on C measured at 59.4~MeV/nucleon \cite{ATB97}.

The numerical conditions of the DEA calculations are similar to those used in Ref.~\cite{GBC06}.

\subsection{Breakup on Pb with the NLO potentials}\label{PbNLO}

\begin{figure}[h]
\includegraphics[width=\linewidth]{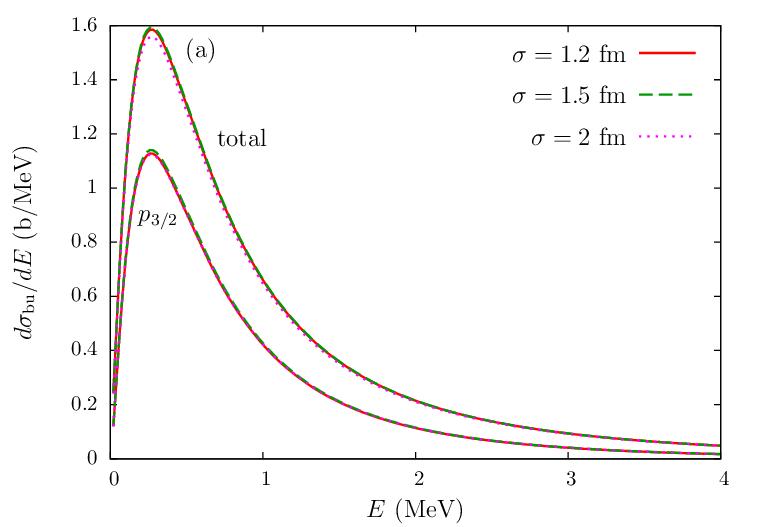}
\includegraphics[width=\linewidth]{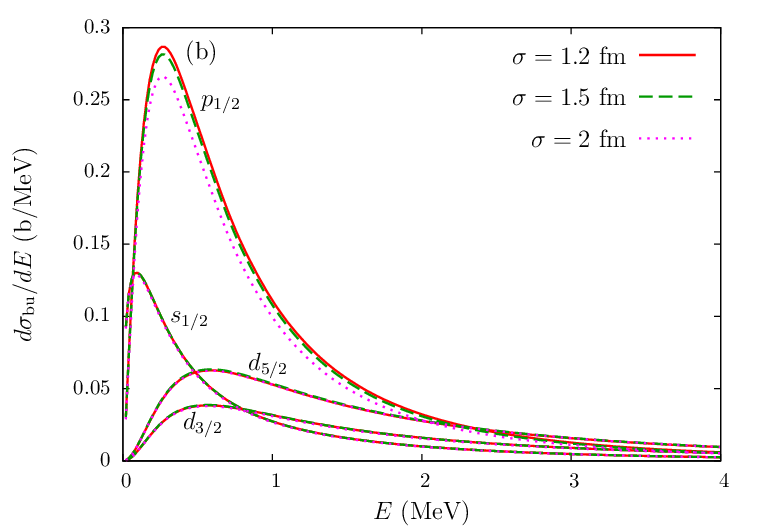}
\caption{\label{f8} Breakup cross section for \ex{11}Be impinging on Pb at 69~MeV/nucleon as a function of the relative energy $E$ between \ex{10}Be and the valence neutron after dissociation.
The calculations are performed with the three NLO Gaussian \ex{10}Be-$n$ potentials described in \Sec{Gaussians}.
(a) The total breakup cross section and the contribution of the dominant $p_{3/2}$ partial wave.
(b) The contribution of the $s_{1/2}$, $p_{1/2}$ and $d$ partial waves.}
\end{figure}

The breakup cross section of \ex{11}Be on Pb at 69~MeV/nucleon is displayed in \Fig{f8} as a function of the energy $E$ between the \ex{10}Be core and the halo neutron after dissociation.
To be able to compare these calculations with the RIKEN data \cite{Fuk04}, this cross section is computed for a \ex{10}Be-$n$ center-of-mass scattering angle $\Theta\le6^\circ$.
The figure presents results obtained from the Gaussian \ex{10}Be-$n$ potentials of ranges $\sigma=1.2$~fm (solid red lines), 1.5~fm (green dashed lines), and 2~fm (magenta dotted lines).
In addition to the total cross section [\Fig{f8}(a)], the contributions of the major partial waves are shown as well: the dominant $p_{3/2}$ in \Fig{f8}(a) and the $s$, $p_{1/2}$ and $d$ ones in \Fig{f8}(b).
It should be remembered that in the Halo-EFT calculations presented in this section the \ex{10}Be-$n$ interaction in the $p_{3/2}$ and higher partial waves is set to zero.
These partial waves still contribute to the cross section---substantially in the case of $p_{3/2}$---even though they are described as plane waves.

All three Halo-EFT \ex{10}Be-$n$ potentials provide nearly identical results, even though they lead to significant differences in the interior of the projectile wave functions (see Figs.~\ref{f1} and \ref{f3}).
This confirms previous studies that have shown breakup reactions to be peripheral \cite{CN07}, in the sense that they probe only the tail of the projectile wave function, and are basically insensitive to its short-range description.
Our new analysis shows that when the asymptotics of the projectile wave functions, viz. their ANC and phase shifts, are constrained, the computed breakup cross section is independent of the interaction used to obtain these nuclear-structure outputs.

Although the three \ex{11}Be descriptions lead to nearly identical breakup cross sections, there is interesting information in this figure pertaining to the small differences between the three calculations.
At the maximum, the $\sigma=1.2$~fm potential leads to a cross section barely lower than the $\sigma=1.5$~fm one, while the $\sigma=2$~fm cross section lies about 10\% lower than the other two.
These differences are attributable mostly to the $p_{1/2}$ contributions [see \Fig{f8}(b)], which scale as the $p_{1/2}$ phase shifts (see \Fig{f4}): those for the $\sigma=1.2$~fm and $\sigma=1.5$~fm potentials are nearly superimposed, whereas that obtained for the broader $\sigma=2$~fm potential lies lower.
The three $s_{1/2}$ contributions on the contrary are nearly identical, as expected from the fact that the corresponding phase shifts are very close to one another (see \Fig{f2}).
This confirms the role played by the continuum description, and hence by the phase shifts, in breakup calculations (see \Sec{eft} and Refs.~\cite{Typel:2004zm,TB05,CN06}).

For all the potentials considered here, the other partial waves are described identically by plane waves.
Therefore, besides the couplings within the continuum, the only difference between them should be due to the ANC of the \ex{11}Be ground state; and this is what we observe.
For example, in the $p_{3/2}$ contribution, the calculation preformed with the $\sigma=1.5$~fm Gaussian potential lies about 1\% higher than the other two; this corresponds exactly to the small difference in ${\cal C}_{1s1/2}$ that can be noted in \tbl{t1}.

Similar results are obtained in other breakup observables measured at RIKEN \cite{Fuk04}, such as the energy distribution obtained with a forward-angle cut ($\Theta<1.3^\circ$) and the angular distribution, where the breakup cross section is integrated in a low-energy range and given as a function of the scattering angle $\Theta$.

\begin{figure}[ht]
\includegraphics[width=\linewidth]{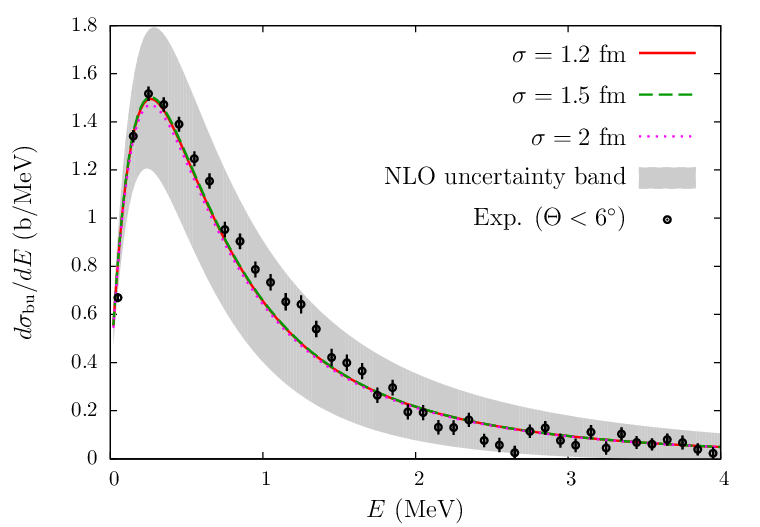}
\caption{\label{f8b} Comparison of the theoretical predictions shown in \Fig{f8} to the data of RIKEN Ref.~\cite{Fuk04} for the breakup of \ex{11}Be impinging on Pb at 69~MeV/nucleon.
The theoretical cross sections have been folded with the experimental energy resolution.}
\end{figure}

To directly compare our results with the RIKEN data \cite{Fuk04}, it is necessary to fold our theoretical calculations with the experimental resolution.
This is done in \Fig{f8b}.
We observe a rather good agreement between our calculations and the data: all three curves would require no or little adjustment in magnitude to exactly fit the data.

One of the advantages of the Halo EFT over usual phenomenological descriptions of halo nuclei is its inherent estimation of the uncertainty within the model.
We estimate the relative error due to the truncation of the EFT expansion to the NLO by $\frac{E+S_{1n}}{E(2^+)+S_{1n}}$, where $E(2^+)$ is the excitation energy of the first $2^+$ state of the \ex{10}Be core, the first degree of freedom neglected in this model of \ex{11}Be.
The corresponding uncertainty is provided as the grey band in \Fig{f8b}.
All the experimental points lie within that uncertainty band, which confirms that the truncated orders and/or missing degrees of freedom most likely explain the remaining discrepancy between the theory and the experiment.

\subsection{Results with LO Halo-EFT potentials}

\begin{figure}[ht]
  \includegraphics[width=\linewidth]{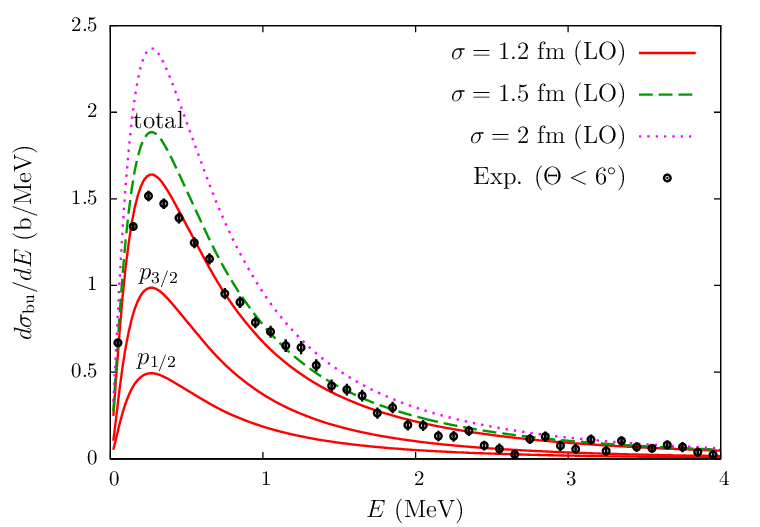}
\caption{\label{f8LO} The significance of the NLO term in the Halo-EFT description of \ex{11}Be is estimated by computing the breakup cross section for \ex{11}Be impinging on Pb at 69~MeV/nucleon using the LO Gaussian \ex{10}Be-$n$ potentials given in \Sec{LO}.
The total breakup cross section for all three potentials together with the contributions of the dominant $p_{3/2}$ and $p_{1/2}$ partial waves for the specific case of the $\sigma=1.2$~fm potential are shown.
The data of Ref.~\cite{Fuk04} are provided for comparison.}
\end{figure}

To evaluate the role played by the NLO term in the Halo-EFT expansion, we repeat the calculations with the LO \ex{10}Be-$n$ potentials described in \Sec{LO}.
The difference with the NLO analysis of the previous section is thus that the ground-state ANC is no longer constrained and that the $p_{1/2}$ wave no longer hosts the $\half^-$ excited bound state of \ex{11}Be.
The results shown in \Fig{f8LO} illustrate that, as expected, the variation of the cross section with the potential width $\sigma$ is markedly larger at LO than at NLO (compare to \Fig{f8}).

The variation between these LO results is due to the change in ANC obtained with different $\sigma$s (see \tbl{t1LO}).
It goes away when the cross sections for the three values of $\sigma$ are all scaled to the same value of the ANC.
Note however that the ANC is not the sole \ex{11}Be-structure observable at play in this reaction.
As already pointed out in \Sec{eft}, the $p_{1/2}$ phase shift plays a non-negligible role.
If this were not the case, the $\sigma=1.5$~fm calculation, for which the ANC remains the same at LO and NLO would be the same in Figs.~\ref{f8} and \ref{f8LO}.
However, it is not the case: constraining the $p_{1/2}$ wave to host the $\half^-$ state of \ex{11}Be puts a constraint on that phase shift (see \Fig{f2}), which reduces the corresponding contribution to the breakup cross section as shown by \Eq{eq:dBdE1NLOp12}---bearing in mind that the effective range $r_{p1/2}$ is negative (see \tbl{t2}).
Therefore, the confrontation of LO calculations to data cannot be used to directly extract an ANC for the \ex{11}Be ground state, as already pointed out in Ref.~\cite{CN17}.

\subsection{Breakup on C with the NLO potentials}

\begin{figure}[ht]
\includegraphics[width=\linewidth]{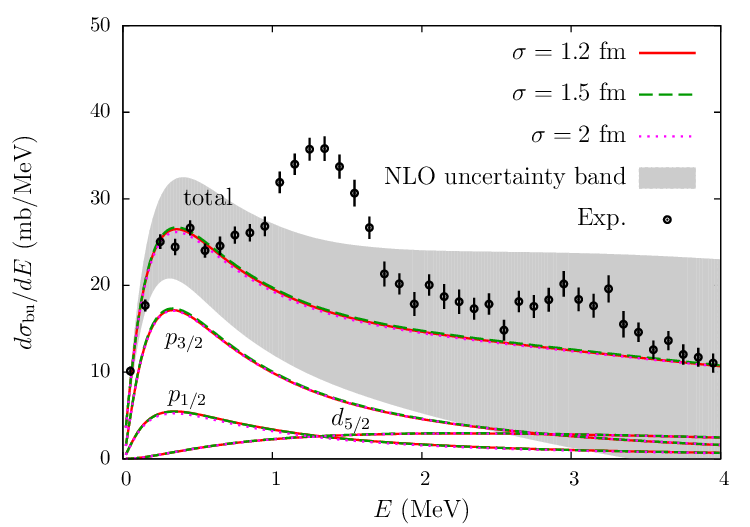}
\caption{\label{f9} Breakup cross section for \ex{11}Be impinging on C at 67~MeV/nucleon as a function of the \ex{10}Be-$n$ energy $E$.
The calculations are performed with the three NLO Gaussian \ex{10}Be-$n$ potentials described in \Sec{Gaussians}.
The contributions of the dominant $p_{3/2}$, $p_{1/2}$ and $d_{5/2}$ partial waves are shown separately.
The data of Ref.~\cite{Fuk04} are shown for comparison.}
\end{figure}

The Coulomb breakup of \ex{11}Be is not the sole measurement that has been reported by Fukuda \etal\ in Ref.~\cite{Fuk04}.
They have also studied the breakup on a carbon target at 67~MeV/nucleon.
Because the DEA has been shown to be valid on both heavy and light targets at these energies, we complete our study by a series of calculations for that collision.
The breakup cross sections obtained on C with the three Gaussian potentials are displayed in \Fig{f9} as a function of the relative energy $E$ between the \ex{10}Be core and the halo neutron after dissociation.
The contributions of the dominant $p_{3/2}$, $p_{1/2}$ and $d_{5/2}$ partial waves are shown separately.

All three NLO \ex{10}Be-$n$ potentials lead to nearly identical cross sections, hence confirming the result of Ref.~\cite{CN07}, where it was seen that, even when the reaction is dominated by nuclear interactions, it remains sensitive mainly to asymptotic properties of the halo-system wave function.
As for the Coulomb breakup, the tiny differences between these calculations can be directly related to the small changes in the ANC or phase shifts produced by the potentials (see \tbl{t1} and \Sec{potentials1}).

Although the calculations have not been convoluted with the experimental energy resolution, we include the RIKEN data in \Fig{f9} to estimate the accuracy of our calculations.
Besides the dominant $p$-wave contributions, we observe that the nuclear interaction between the projectile and the target significantly populates higher partial waves, like the $d_{5/2}$ illustrated here.
This interaction also leads to a slower decrease of the cross section with the continuum energy $E$.
These two effects combine to provide a total cross section that is flatter than in Coulomb breakup.
This behavior is in very good agreement with the data.
However, although they reproduce most of the experimental cross section, our calculations miss its structure above $E=1$~MeV, viz. the two bumps at about 1.3~MeV and around 3~MeV.
As explained in Refs.~\cite{Fuk04,CGB04}, these structures correspond to the $\frac{5}{2}^+$ and $\frac{3}{2}^+$ resonant states in the \ex{11}Be continuum spectrum.
In an extreme shell model of the nucleus, they correspond to $d_{5/2}$ and $d_{3/2}$ single-neutron resonances.
They do not appear in the theoretical cross section because they are not included in our NLO description of the projectile.
In the next two sections, we explore this part of the model space by extending the (strict) Halo-EFT expansion to the next three partial waves: $d_{5/2}$, $p_{3/2}$, and $d_{3/2}$, for which experimentally-known resonant states can be used to adjust our Gaussian potentials.

\section{\label{potentials2} Adding resonances to the NLO Halo EFT potential}

We now turn our attention to what is needed for an improved description in waves other than the $s_{1/2}$ and $p_{1/2}$. At this point we leave the strict Halo-EFT expansion of Ref.~\cite{Hammer:2011ye} and embark on a more phenomenological investigation, in an attempt to find the ingredients that are needed to describe the data at energies beyond those where the EFT expansion breaks down. 

\subsection{\label{d5}$d_{5/2}$ wave}

\begin{figure}[ht]
\includegraphics[width=\linewidth]{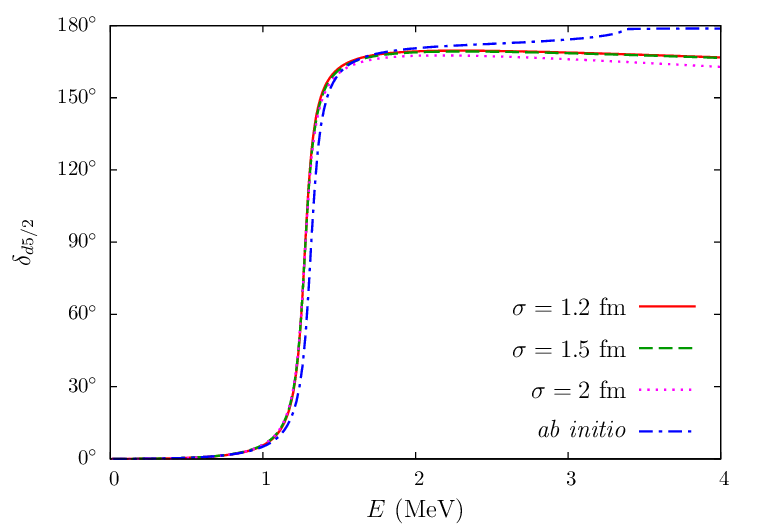}
\caption{\label{f6} Phase shifts obtained in the $d_{5/2}$ partial wave from our effective single-particle description of \ex{11}Be and the NCSMC calculation of Calci et al. \cite{CNR16}.}
\end{figure}

There is no physical bound state to which we could fit our Gaussian potentials in the $d_{5/2}$ partial wave.
However, there is a narrow $\fial^+$ state in the low-energy \ex{10}Be-$n$ continuum, which is usually seen as a single-particle $d_{5/2}$ resonance: $E_{d5/2}=1.274$~MeV with $\Gamma_{d5/2}=100$~keV \cite{CGB04}\footnote{These values differ slightly from the more recent Ref.~\cite{KKP12} for the same reason mentioned in \Sec{eft}.}.
This seems confirmed by the structure observed in the $d_{5/2}$ phase shift computed within the NCSMC framework of Calci \etal\ (see Fig.~3 of Ref.~\cite{CNR16} and the blue dash-dotted line in \Fig{f6}).
To constrain our Gaussian potentials in this partial wave, we choose to reproduce the experimental energy and single-particle width of this resonant state. (The NCSMC calculation produces a resonance that is slightly displaced from the experimental location, and we choose to reproduce the latter.)
The depths of the Gaussian potentials obtained in this fit, and the energy and width of the $d_{5/2}$ resonance they produce, are listed in \tbl{t4} alongside the \ai\ results of Ref.~\cite{CNR16}.
The corresponding phase shifts are plotted in \Fig{f6}.

As in the $s_{1/2}$ wave, all potentials provide very similar phase shifts, which also agree very well with the \ai\ prediction up to $E=2$~MeV. 

\begin{table}
\begin{tabular}{c|cc|cc}\hline\hline
$\sigma$ & $V^{(0)}_{d5/2}$ & $V^{(2)}_{d5/2}$ & $E_{d5/2}$ & $\Gamma_{d5/2}$\\ 
(fm) & (MeV) & (MeV fm\ex{-2}) & (MeV) &  (MeV)\\\hline
1.2 & -106.61 & -30 & 1.274 & 0.096\\
1.5 & -139.945 & +2 & 1.274 & 0.101\\
2.0 & -104.597 & +4.5 & 1.274 & 0.105\\\hline
\ai & - & -& 1.31 & 0.1\\
\hline\hline
\end{tabular}
\caption{\label{t4} Depths of the Gaussian \ex{10}Be-$n$ potentials \eq{e5} obtained by fitting the position and width of the $d_{5/2}$ resonance in
\ex{11}Be~\cite{KKP12}. The energy and width of this resonance found in the \ai\  calculations of Calci \etal\ \cite{CNR16} are also provided.}
\end{table}

\subsection{\label{p3}$p_{3/2}$ wave}

\begin{figure}[ht]
\includegraphics[width=\linewidth]{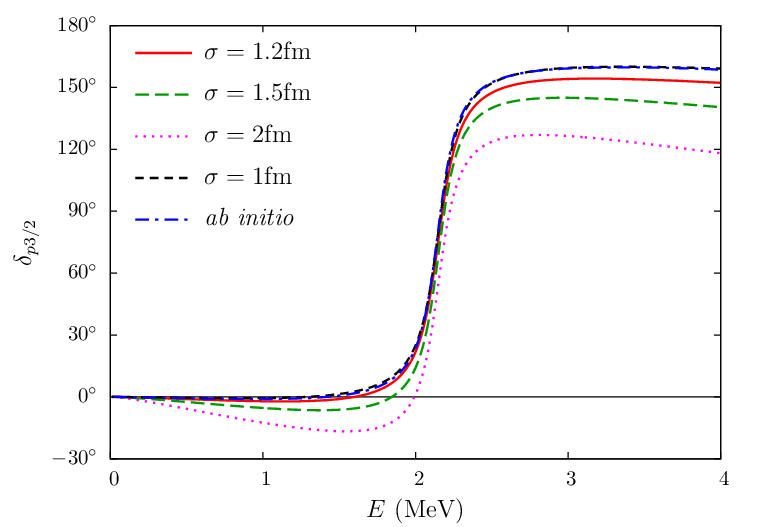}
\caption{\label{f5} Phase shifts obtained in the $p_{3/2}$ partial wave with the three usual Gaussian potentials ($\sigma=1.2$, 1.5, and 2.0~fm).
The prediction of Calci \etal\ \cite{CNR16} is shown, as well as the best fit obtained with a narrow Gaussian potential ($\sigma=1.0$~fm).}
\end{figure}

In the $p_{3/2}$ partial wave, we fit the depth of the potentials to reproduce the experimental energy and width of the $\thal^-$ resonant state: $E_{3/2^-}=2.15$~MeV and $\Gamma_{3/2^-}=210$~keV \cite{KKP12}.
The resulting potential depths are provided in the first three lines of \tbl{t3} and the corresponding phase shifts are plotted in \Fig{f5} together with Calci \etal's results, which exhibit the clear signature of a single-particle resonance.
Interestingly, the \ai\ phase shift is consistent with 0 below $E\sim 1.5$~MeV, which confirms our choice made in the NLO description to neglect the $^{10}$Be-$n$ interaction in that partial wave.

Although they have been adjusted to the same resonance properties, the three Gaussian potentials lead to different phase shifts, suggesting that some short-range physics plays a role in the structure of this state and that a simple single-particle description will not be enough to describe it accurately.
Moreover, only the narrow potential ($\sigma=1.2$~fm) yields phase shifts in good agreement with the \ai\ prediction.
If, as we expect, this phase shift plays a role in the breakup reaction, we should expect some differences between all three calculations, at least in the $p_{3/2}$ contribution to the breakup cross section \cite{CN06}.

Since our goal is to reproduce the output of the NCSMC calculation of Calci \etal, we develop, for this partial wave, a potential with $\sigma=1$~fm, which reproduces nearly perfectly the \ai\ $p_{3/2}$ phase shift (see the back dashed line in \Fig{f5}; the depth of this potential is provided in the fourth line of \tbl{t3}).
Since it best simulates the \ai\ predictions in this partial wave, this potential is probably the one that should be considered as the best effective description of \ex{11}Be to be used in reaction models.

\begin{table}
\begin{tabular}{c|cccc}\hline\hline
$\sigma$ & $V^{(0)}_{p3/2}$ & $V^{(2)}_{p3/2}$ & $E_{p3/2}$ & $\Gamma_{p3/2}$\\ 
(fm) & (MeV) & (MeV fm\ex{-2}) & (MeV) &  (MeV)\\\hline
1.2 & -599.545 & +138 & 2.150 & 0.211\\
1.5 & -386.112 & +58 & 2.150 & 0.203\\
2.0 & -215.84 & +18.5 & 2.150 & 0.209\\\hline
1.0 & -860.292 & +282 & 2.150 & 0.207\\\hline
\ai & - & -& 2.15 & 0.19\\
\hline\hline
\end{tabular}
\caption{\label{t3} Depths of the Gaussian \ex{10}Be-$n$ potentials \eq{e5} obtained by fitting the resonance position and width from Ref.~\cite{KKP12} in the $\thal^-$ partial wave.
The energy and width of the $p_{3/2}$ resonance obtained in the \ai\  calculation of Calci \etal\ \cite{CNR16} are also shown.}
\end{table}

\subsection{\label{d3}$d_{3/2}$ wave}

\begin{figure}[ht]
\includegraphics[width=\linewidth]{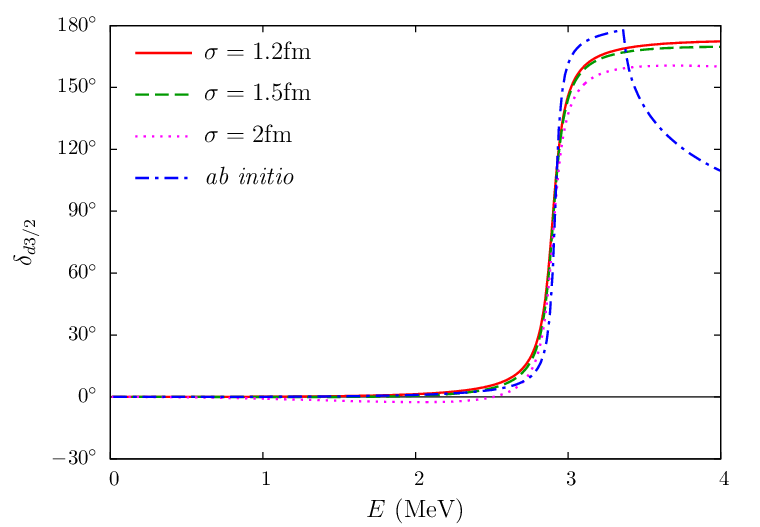}
\caption{\label{f7} Phase shifts obtained in the $d_{3/2}$ partial wave, including the \ai\ calculation of Calci et al. \cite{CNR16}.}
\end{figure}

The depths of the Gaussian potentials in the $d_{3/2}$ partial wave are adjusted to reproduce the energy and width of the $\thal^+$ resonant state: $E_{3/2^+}=2.90$~MeV and $\Gamma_{3/2^+}=122$~keV \cite{KKP12}.
These depths are listed in \tbl{t5} and the resulting phase shifts are displayed in \Fig{f7}.
Albeit better than in the $p_{3/2}$ partial wave, these fits are not very satisfactory.
First, the different potentials lead to significant differences in the phase shifts.
Second, they do not reproduce the \ai\ calculation very accurately.
This is partly due to the fact that the NCSMC predicts too narrow a $d_{3/2}$ resonance: Calci \etal\ obtain $\Gamma_{d3/2}=60$~keV, instead of the experimental 122~keV.

It is usually believed that this state is not well described by a single-particle model in which the \ex{10}Be core is in its $0^+$ ground state. Instead, a significant part of the overlap wave function corresponds to a configuration in which the core is in its first $2^+$ state \cite{ML12}.
The influence of that state is visible in the cusp observed in the \ai\ phase shift at $E\approx3.4$~MeV.
We therefore do not expect our effective description of the $\thal^+$ state of \ex{11}Be to fully account for the underlying physics of that state.

\begin{table}
\begin{tabular}{c|cccccc}\hline\hline
$\sigma$ & $V^{(0)}_{d3/2}$ & $V^{(2)}_{d3/2}$ & $E_{d3/2}$ & $\Gamma_{d3/2}$\\ 
(fm) & (MeV) & (MeV fm\ex{-2}) & (MeV) &  (MeV)\\\hline
1.2 & -325.87 & +50 & 2.90 & 0.12\\
1.5 & -229.22 & +28 & 2.90 & 0.117\\
2.0 & -136.98 & +11 & 2.90 & 0.122\\\hline
\ai & - & -& 2.92 & 0.06\\
\hline\hline
\end{tabular}
\caption{\label{t5} Depths of the Gaussian \ex{10}Be-$n$ potentials \eq{e5} obtained by fitting resonance position and width from Ref.~\cite{KKP12} in the $d_{3/2}$ partial wave.
The energy and width of the $d_{3/2}$ resonance obtained in the \ai\ calcuation of Calci \etal \cite{CNR16} are also provided.}
\end{table}

\section{\label{results2} Results with NLO + resonances Halo EFT potential}
\subsection{Coulomb breakup}

We now augment the previous calculation of Coulomb breakup on \ex{208}Pb at 69~MeV/nucleon by including the resonances in the $d_{5/2}$, $p_{3/2}$, and $d_{3/2}$ channels, using the potentials described in the previous section. All other ingredients of the reaction calculation---viz. the descriptions of the $s_{1/2}$ and $p_{1/2}$ partial waves, the target-neutron and target-\ex{10}Be potentials, and the numerical parameters---are as in Sec.~\ref{results1}. 

\begin{figure}[ht]
\includegraphics[width=\linewidth]{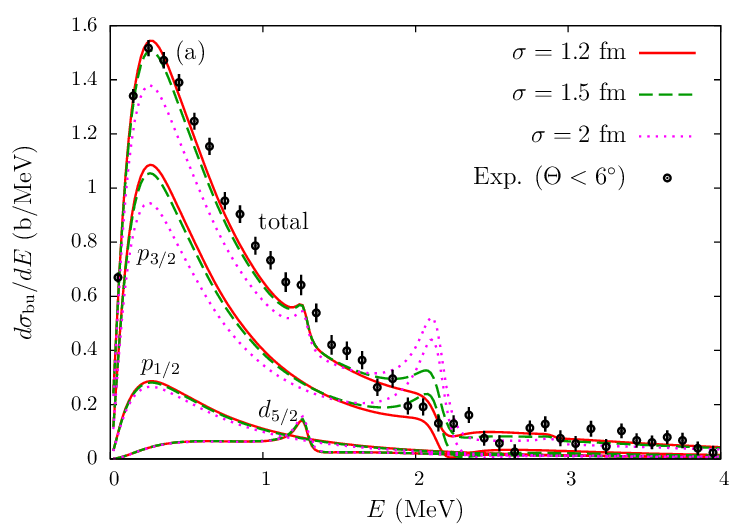}
\includegraphics[width=\linewidth]{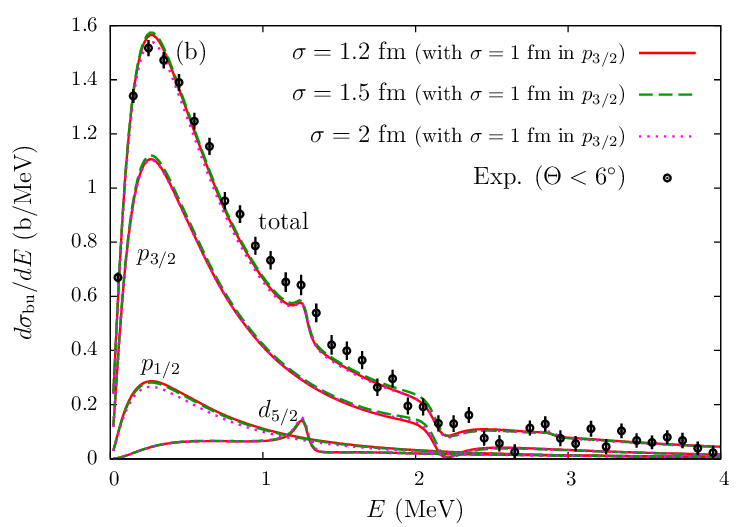}
\caption{\label{f10} 
Study of the role played by the resonant continuum in the breakup cross section for \ex{11}Be on Pb at 69~MeV/nucleon.
The calculations performed with the three Gaussian \ex{10}Be-$n$ potentials fitted in \Sec{potentials2} to reproduce the $\fial^+$, $\thal^-$, and $\thal^+$ resonant states are displayed in the upper panel (a).
In the lower panel (b), the cross sections are obtained with the same potentials, but in the $p_{3/2}$ partial wave we use the narrow Gaussian potential ($\sigma=1$~fm), which best reproduces the \ai\ $\delta_{3/2^-}$.
In both panels the contributions of the dominant $p_{3/2}$, $p_{1/2}$, and $d_{5/2}$ partial waves are shown separately.
Although no folding with the experimental resolution has been performed, the experimental data of Ref.~\cite{Fuk04} are shown for comparison.}
\end{figure}

The corresponding DEA cross sections are displayed in \Fig{f10}.
The upper panel (a) presents the results obtained with the three Gaussian potentials of widths $\sigma=1.2$, 1.5, and 2~fm detailed in \Sec{potentials2}.
Contrary to the previous series of tests, we note significant differences between these potentials: there is a significant variation in the magnitude at the maximum of the cross section and the behavior around the $\thal^-$ resonance changes a lot from one potential to the other.
As expected, these differences are mostly due to the $p_{3/2}$ contribution and can be directly related to the differences in the $p_{3/2}$ phase shifts obtained with the three potentials (see \Fig{f5}).

In the vicinity of the maximum, the ordering of the curves follows that of the phase shifts: the $\sigma=2$~fm potential, producing the most negative phase shift, gives the lowest $p_{3/2}$ contribution to the breakup cross section, while the $\sigma=1.2$~fm potential, generating the phase shift closest to zero, leads to the largest cross section.

Near the $\thal^-$ resonance, the strikingly different behavior between the potentials can be  related to the way that state is described.
The $\sigma=2$~fm potential, which produces a $p_{3/2}$ phase shift at odds with the \ai\ prediction, leads to a large bump in the cross section.
Since this bump is not seen in the data, we can infer that this description of the $p_{3/2}$ phase shift is irrealistic.

These results confirm the analysis of Ref.~\cite{CN06}, which showed that breakup calculations are affected by variations in the description of the continuum of the projectile, viz. by the phase shifts.
This is the reason why we have developed a fourth Gaussian potential with a narrow width $\sigma=1$~fm that produces a phase shift in nearly perfect agreement with the NCSMC one (see the black short dashed line in \Fig{f5}).
We have repeated our calculations using that potential in the $p_{3/2}$ partial wave while keeping the Gaussian potentials with the different widths ($\sigma=1.2$, 1.5, and 2~fm) in the other partial waves.
The resulting cross sections are displayed in \Fig{f10}(b).

Once the $p_{3/2}$ partial wave is fixed in this way, the calculations are nearly independent of $\sigma$, confirming the results of \Sec{results1}, where it was suggested that the reaction is purely peripheral and is not affected by the short-range physics of the projectile.
At low energy---viz. $E\lesssim 1$~MeV---they are nearly identical to those obtained at the strict NLO level (see \Fig{f8}), at which the partial waves other than $s_{1/2}$ and $p_{1/2}$ are described by plane waves.
This is is due to the nearly zero $p_{3/2}$ phase shift obtained with the narrow $\sigma=1$~fm Gaussian potential in that energy range.

At the $\fial^+$ resonance energy, we observe a small bump in the $d_{5/2}$ contribution, which seems to fit the bump observed in the data at the same energy.
Unfortunately, this bump is washed out by the convolution with the experimental resolution, which suggests that the description of that $\fial^+$ state is probably more complex than the single-particle description considered here.

The breakup cross section obtained with all three potentials now exhibits the same behavior in the vicinity of the $\thal^-$ resonance.
Although that particular shape disappears after convolution with the experimental resolution, let us note that it is very similar to the energy dependence of the \ai\ $\frac{d{\rm B(E1)}}{dE}$ computed by Calci \etal\ (see Fig.~5 of Ref.~\cite{CNR16}).

At larger energy, the effect of the $\thal^+$ state on the cross section is marginal.
Since it does not seem to affect the experimental data, we cannot conclude anything about the accuracy of the Halo-EFT description of this state in the $d_{3/2}$ partial wave.

We thus observe that, at least for Coulomb breakup, the  nuclear-structure properties that ultimately matter in the description of reactions involving halo nuclei are the ANC of the ground state and the phase shifts in the dominant partial waves.
Interestingly, Halo-EFT provides a useful organization of the set of these properties that are important for low energy.
Once the $s_{1/2}$ and $p_{1/2}$ potentials are fixed from the \ai\ results a $\sigma$-independent description of the breakup cross section that agrees with the experimental data is found.
We note that in the strict NLO Halo EFT calculation the $p_{3/2}$ wave is treated as free, and in fact this produces a better outcome for the breakup cross section than is obtained with a broad $p_{3/2}$ potential tuned to the resonance energy.
Unless the $p_{3/2}$ partial wave is tuned to agree with \ai\ input simply setting this interaction to zero may produce a better result than does an inclusion of the $p_{3/2}$ resonance which results in unconstrained low-energy $p_{3/2}$ phase shifts.

\subsection{\label{nucleartarget}Nuclear breakup}

The Coulomb-breakup calculations presented in the previous section are not strongly sensitive to the resonances added phenomenologically to the NLO description of $^{11}$Be.
As explained at the end of \Sec{results1}, the nuclear breakup on $^{12}$C is much more affected by these resonances.
In this section, we study how the single-particle description of these resonances detailed in \Sec{potentials2} affects the breakup calculations and how they compare to experiment.
As on Pb, we consider all three \ex{10}Be-$n$ Gaussian potentials.
However, following the conclusion of the previous section, we have systematically used the $\sigma=1$~fm potential in the $p_{3/2}$ partial wave, which best reproduces the NCSMC $\thal^-$ phase shift.

\begin{figure}[ht]
\includegraphics[width=\linewidth]{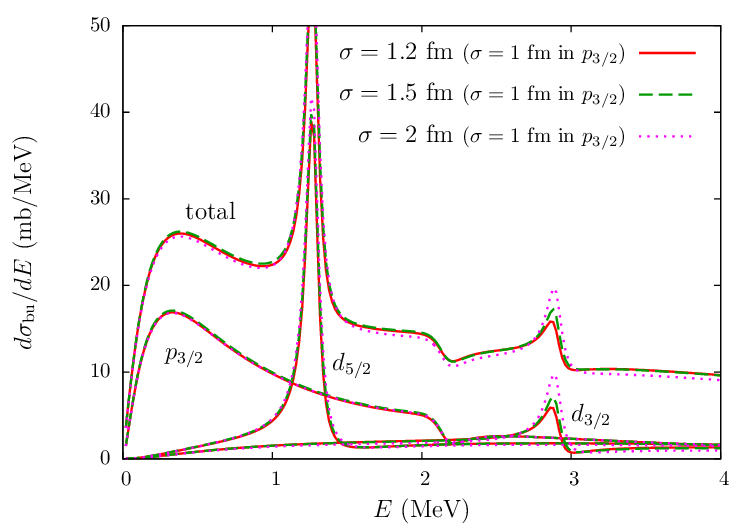}
\caption{\label{f11}
Breakup cross section for \ex{11}Be impinging on C at 67~MeV/nucleon as a function of the \ex{10}Be-$n$ energy $E$.
The calculations are performed with the three Gaussian \ex{10}Be-$n$ potentials described in Secs~\ref{Gaussians} and \ref{potentials2} assuming a $\sigma=1.0$~fm in the $p_{3/2}$ partial wave following the constraint obtained from our analysis of the Coulomb breakup.
The contributions of the dominant $p_{3/2}$, $d_{5/2}$ and $d_{3/2}$ partial waves are shown separately.}
\end{figure}

The results of these calculations are displayed in \Fig{f11}.
The total breakup cross section obtained from our dynamical calculations as well as the $p_{3/2}$, $d_{5/2}$, and $d_{3/2}$ main contributions are plotted as a function of the \ex{10}Be-$n$ relative energy $E$.
As already seen in Ref.~\cite{CGB04}, the contributions of the $d$ waves are no longer negligible and dominate the calculation in the vicinity of the resonances.
This is especially true for the $d_{5/2}$ partial wave, whose contribution exhibits a huge and narrow peak at the resonance energy.
That peak has the same width as the resonance ($\Gamma_{d5/2}\approx100$~keV), which confirms that nuclear breakup is, in principle, an excellent tool to study unbound states in loosely-bound nuclei \cite{Fuk04,CGB04}.
Albeit as narrow as the $d_{5/2}$ continuum state, the $d_{3/2}$ resonance produces a much smaller peak in the cross section.

With the potential constrained in the $p_{3/2}$ partial wave, all three calculations provide very similar cross sections, independently of the potential range $\sigma$.
Even here, in the collision with \ex{12}C, there is not great sensitivity to the interior of the \ex{11}Be wave function.
In particular, the three $d_{5/2}$ contributions are nearly superimposed on each other.
These three potentials describe the \ai\ $d_{5/2}$ phase shift equally well (see \Fig{f6}) but do differ in what they predict for the interior of the wave function in that channel. 
On the contrary, in the $d_{3/2}$ partial wave, the three potentials lead to different peaks in the second resonance region.
However, once again, these differences between potentials can be directly related to their different phase shifts (see \Fig{f7}).
These results confirm the sensitivity of breakup calculations to the description of the continuum of the projectile.
In particular,  similar descriptions of the phase shifts will lead to similar partial-wave contributions to the breakup cross section, independent of the details of the potential used to generate the continuum states.

\begin{figure}[ht]
\includegraphics[width=\linewidth]{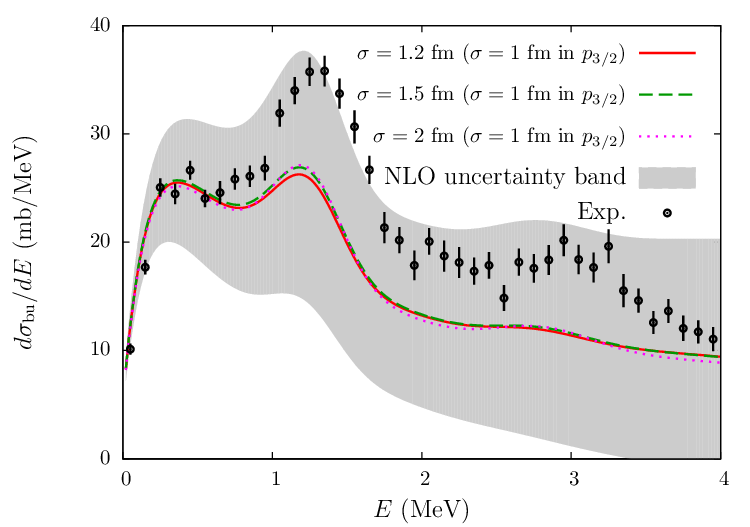}
\caption{\label{f12} 
Comparison of the theoretical prediction for the breakup cross section for \ex{11}Be impinging on C at 67~MeV/nucleon with the experimental data of Ref.~\cite{Fuk04}.
Our calculations have been convoluted with the experimental resolution.}
\end{figure}

To confront these results with the RIKEN data, it is necessary to fold our calculations with the energy resolution of the experiment, see \Fig{f12}.
As already observed in Ref.~\cite{CGB04}, this comparison shows that a simple single-particle description of the projectile reproduces most of the breakup strength, providing a theoretical cross section rather close to the experimental value.
However, it also shows that the breakup strength in both $d$ resonances is clearly underestimated in our single-particle model of the projectile.
As already suggested in the calculations on the Pb target, our description of the $d_{5/2}$ does not provide the entire breakup strength measured experimentally.
And the small peak induced by the single-particle $d_{3/2}$ resonance is completely washed out by the folding, indicating that the $\thal^+$ resonance cannot be reliably described within a single-particle model of \ex{11}Be.
We emphasize that these deficiencies in our reaction cross section are almost certainly not due to the simple potentials used.
If the results were sensitive to the form of the potential then we would see a $\sigma$ dependence in the cross sections predicted here.

Even though the data are covered by the NLO uncertainty band, which suggest that they could be reached by the calculation, were they performed with higher orders in the Halo-EFT expansion, these results suggest that our calculation is missing a degree (or degrees) of freedom.
In this case the lack of strength in the $d$ resonances seems to confirm the work of Moro and Lay, who have shown within a DWBA model the significant role played by the first $2^+$ excited state of the core in these resonant breakup reactions \cite{ML12}.

\section{\label{unobservable}Spectroscopic factor vs. ANC}

\begin{figure*}[t]
\includegraphics[width=0.49\linewidth]{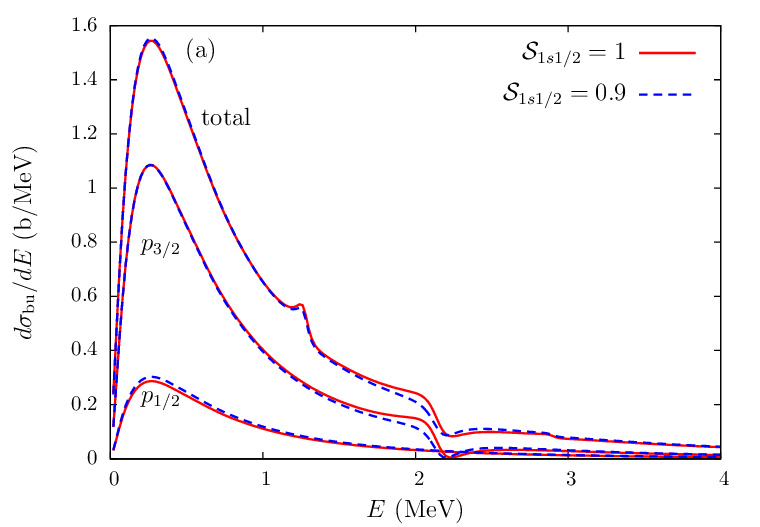}
\includegraphics[width=0.49\linewidth]{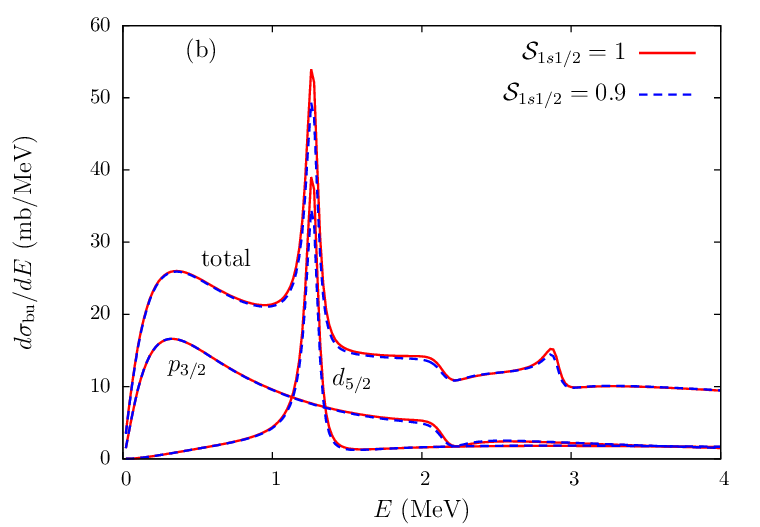}
\caption{\label{f14} Difference between breakup calculations with a \ex{11}Be bound-state wave function normalized to unity or to the spectroscopic factor predicted by the \ai\ calculation \cite{CNR16}. Left on Pb at 69~MeV/nucleon. Right on C at 67~MeV/nucleon.}
\end{figure*}

So far all breakup calculations have been performed using single-particle wave functions normalized to  unity, even though the NCSMC calculations of Calci \etal\ predict a spectroscopic factor ${\cal S}_{1s1/2}=0.90$ for the $s$-wave configuration of the ground state \cite{CNR16}.
The analysis presented in Ref.~\cite{CN07} shows that breakup reactions, both on light and heavy targets, are purely peripheral and hence that no information about possible spectroscopic factors could be extracted from them.
In this section, we test this result by repeating our calculations with an initial ground-state wave function normalized to the spectroscopic factor predicted by the \ai\ model.
To this end, we have refitted the Gaussian potential in the $s$ wave to conserve the same ANC while reducing the norm of the wave function to $\sqrt{0.90}$.
Since we could not find a combination of depths that would allow this for the $\sigma=1.2$~fm potential, we have used a slightly larger range of the Gaussian form factor ($\sigma=1.3$~fm).
(This is related to the Wigner bound discussed in Sec.~\ref{sec:iteration}.)
In \Fig{f14} we confront the corresponding results (blue dashed lines) with the calculation performed with a wave function normalized to one (red solid line) for both the Pb (a) and C (b) targets.

For both Coulomb- and nuclear-dominated breakups, we observe virtually no difference between the two calculations, hence confirming the results of Ref.~\cite{CN07}.
This means that---at least in breakup reactions---spectroscopic factors are not observables.
Interestingly, the only energies where the reaction seems slightly more sensitive to the internal part of the wave function are in the vicinity of the resonances.
As already noted in Ref.~\cite{CN07}, these are the only places where both calculations are notably different.
The present work provides a more quantitative estimate of this effect.
A reduction of 10~\% in the spectroscopic factor while keeping the same ANC leads to a reduction of 10--15~\% in the contribution of the resonant partial wave at the resonance energy.
This indicates that resonant-breakup reactions are more affected by the internal part of the projectile wave function and hence could be more sensitive to short-range physics like couplings with other configurations, as suggested by the work of Moro and Lay \cite{ML12}.

\section{\label{conclusion}Conclusions and prospects}

In this work, we have successfully coupled a Halo-EFT description of $^{11}$Be to the Dynamical Eikonal Approximation.
This has enabled us to include an efficient effective description of halo nuclei guided by the principles of halo EFT within an accurate and fully dynamical model of nuclear reactions. 
Our description includes the dominant nuclear-structure inputs, like the ANC of bound states and phase shifts in the continuum, in the reaction modeling.

This approach works well for peripheral reactions.
A detailed description of the projectile's interior is not necessary: we just need to reproduce the asymptotic (and hence observable) part of the wave function, both in the bound state and in the continuum region.
In particular, we have explicitly demonstrated that our results are independent of non-observable quantities: details of the potential at short distances and spectroscopic factors. Our results hence provide another example of the fact that the neutron-core potential is not observable.
Instead, the input information for the halo EFT description can be taken from observables measured in experiment or calculated \ai.
For our example, \ex{11}Be, we have matched the effective description of the projectile to a recent \ai\ calculation in No-Core Shell Model with Continuum (NCSMC)~\cite{CNR16}. 

Halo EFT organizes the \ex{10}Be-neutron properties needed for a description of a given fidelity: here we considered \ex{11}Be models at LO, NLO, and ``NLO-plus-resonances" accuracy. At NLO the inputs needed to construct \ex{10}Be-neutron potentials in the $s_{1/2}$ and $p_{1/2}$ channels are taken from the \ai\
 results of Ref.~\cite{CNR16}. This produces an accurate description of \ex{11}Be collisions with \ex{208}Pb up to about 2 MeV. 
 It also yields a good description of collisions of \ex{11}Be with \ex{12}C up to $E \approx 1$ MeV~\cite{Fuk04},. In all cases we find very little dependence on the range of the Gaussians in EFT results.

At higher energies the nuclear dominated break-up reaction, corresponding to the 
C target, shows some sensitivity to shorter range parts of the
wave function. The inclusion of the \ex{11}Be resonance in the $d_{5/2}$ partial wave produces a feature in the \ex{12}C-\ex{11}Be cross section in the right location, but not one that is strong enough to reproduce the data.  This failure to reproduce the data indicates that relevant degrees of freedom are missing. 
The effect of core excitation to the $2^+$ state in \ex{10}Be can be included as an explicit degree of freedom in halo EFT, although this comes at the expense of additional parameters. 
Such effects, together with the role of neutron-core-target ``three-body" forces, and the limits of  our approach for nuclear dominated reactions also seem fruitful avenues for future work.
Finally, it would be interesting to apply this framework to other types of reactions, such as transfer reactions or reactions of astrophysical interest.
A first study in this direction was carried out in Ref.~\cite{Yang:2018xae,Yang:2018nzr}.

\begin{acknowledgments}
We thank A.~Calci, P.~Navr\`atil, and R.~Roth for interesting discussions and for sending us their numerical phase shifts.
We acknowledge partial support from the ECT* workshops ``Three-body systems in reactions with rare isotopes" and ``Open quantum systems" 
where the present work was initiated and carried forward. We thank the Institute for Nuclear Theory at the University of Washington for support
as part of INT program ``Toward Predictive Theories of Nuclear Reactions Across the Isotopic Chart",
and the Department of Energy for partial support during the completion of this work.
This project has received funding from the European Union's Horizon 2020 research and innovation programme under grant agreement No 654002.
This research was also supported by
the U.S. Department of Energy under grant DE-FG02-93ER40756, 
by the German Federal Ministry of Education and Research under contract 05P15RDFN1, 
by the ExtreMe Matter Institute EMMI at the GSI Helmholtzzentrum f\"ur Schwerionenphysik, Darmstadt, Germany,
by the Deutsche Forschungsgemeinschaft within the Collaborative Research Centers 1044 and 1245,
and the PRISMA (Precision Physics, Fundamental Interactions and Structure of Matter) Cluster of Excellence.
P.~C. acknowledges the support of the State of Rhineland-Palatinate.

\end{acknowledgments}

%

\end{document}